\documentclass[iop]{emulateapj}

\newcommand{\kms}{{\rm km~s\ensuremath{^{-1}}}}
\newcommand{\m}{$M_\bullet$}

\usepackage{natbib}
\usepackage{multirow}
\usepackage{graphicx}

\slugcomment{AJ Accepted}

\shorttitle{A STIS Atlas of Galactic Nuclei}
\shortauthors{Batcheldor et al.}

\begin{document}

\title{A STIS Atlas of CaII Triplet Absorption Line Kinematics in Galactic Nuclei*}
\thanks{*Based on observations made with the NASA/ESA 
Hubble Space Telescope, obtained at the Space Telescope Science Institute, 
which is operated by the Association of Universities for Research in Astronomy, 
Inc., under NASA contract NAS5-26555.}




\author{D. Batcheldor$^1$, D. Axon$^2$, M. Valluri$^3$, J. Mandalou$^1$ and D. Merritt$^4$\\
{\scriptsize $^1$P\lowercase{hysics and} Space Sciences Department, Florida Institute of Technology, 150 West University Boulevard,\\
Melbourne, FL 32901, USA; dbatcheldor@fit.edu\\
$^2$ School of Mathematical and Physical Sciences, University of Sussex, Sussex House, Brighton, BN1 9RH, UK\\
$^3$ Department of Astronomy, University of Michigan, Ann Arbor, MI 48109, USA\\
$^4$ Department of Physics, Rochester Institute of Technology, 84 Lomb Memorial Drive, Rochester, NY 14623-5603, USA}}

\begin{abstract}
The relations observed between supermassive black holes and their host galaxies suggest a fundamental link in the processes that cause these two objects to evolve. A more comprehensive understanding of these relations could be gained by increasing the number of supermassive black hole mass (\m) measurements. This can be achieved, in part, by continuing to model the stellar dynamics at the centers of galactic bulges using data of the highest possible spatial resolution. Consequently, we present here an atlas of galaxies in the Space Telescope Imaging Spectrograph (STIS) data archive that may have spectra suitable for new \m\ estimates. Archived STIS G750M data for all non-barred galactic bulges are co-aligned and combined, where appropriate, and the radial signal-to-noise ratios calculated. The line-of-sight velocity distributions from the CaII triplet are then determined using a maximum penalized likelihood method. We find 19 out of 42 galaxies may provide useful new \m\ estimates since they are found to have data that is comparable in quality with data that has been used in the past to estimate \m.  However, we find no relation between the signal-to-noise ratio in the previously analyzed spectra and the uncertainties of the black hole masses derived from the spectra. We also find that there is a very limited number of appropriately observed stellar templates in the archive from which to estimate the effects of template mismatching. 
\end{abstract}

\keywords{catalogs - galaxies: bulges - stars: kinematics and dynamics} 

\section{Introduction}

The study of supermassive black holes (SBHs) has been one of the most successful areas of Hubble Space Telescope ({\it HST}) research. The importance of SBHs was highlighted with the discovery that their masses (\m) are seen to correlate with many properties of the galaxy bulges that host them, e.g., luminosity \citep{1995ARA&A..33..581K}, stellar velocity dispersion, $\sigma_\ast$ \citep{2000ApJ...539L...9F,2000ApJ...539L..13G}, concentration index \citep{2001ApJ...563L..11G}, and potentially the dark matter halo mass \citep{2002ApJ...578...90F,2003MNRAS.341L..44B}. As these SBH-galaxy scaling relations intimately link the most basic characteristic of a SBH with the defining properties of the surrounding galaxy, they sparked significant efforts toward understanding the nature of black hole and galaxy formation and evolution \citep[e.g.,][]{2001ApJ...552L..13C,2003ApJ...591..125A,2004ApJ...613..109H,2005ApJ...634..910W,2005MNRAS.364..407C,2006ApJ...641...90R,2007ApJ...667..117T,2008arXiv0808.1349C}. 

While the fundamental importance of the SBH scaling relations is clear, their form remains dependent on establishing an unbiased and statistically significant number of high quality direct measurements of \m. However, there are several fundamental difficulties with obtaining such a sample of \m. For example, the relations are currently established over only a few orders of magnitude ($7\lesssim\log$\m$\lesssim9$), there are potential selection effects and biases \citep{2007ApJ...670..249L,2010ApJ...711L.108B}, and there is an increasing population of galaxies that do not seem to follow the same relations defined by nearby early-type galaxies \citep{2009ApJ...698..812G,2010ApJ...721...26G,2011arXiv1102.0537M,2011MNRAS.412.2211G}. In addition, it is not clear whether the local scaling relations are consistent with each other \citep{2007ApJ...663...53T}, nor whether they may have experienced cosmic evolution \citep{2007ApJ...667..117T}. In addition, at present the only way to estimate \m\ outside of the local Universe, and therefore determine if scaling relations have evolved, is to use the reverberation mapping technique \citep[e.g.][]{1993PASP..105..247P} or derivatives thereof \citep{1999ApJ...526..579W}. However, since about 2004, the geometrical factor in the reverberation-mapping virial relation has been calibrated using the \m$-\sigma_\ast$ relation \citep{2004ApJ...615..645O}, increasing still more the importance of establishing this relation.

\begin{deluxetable*}{lccccccccc}
\vspace{-0.5cm}
\tablewidth{510pt}
\tabletypesize{\footnotesize}
\tablecaption{Available STIS CaII Stellar Absorption Line Data for Standard Stars\label{tab:standards}}
\tablewidth{0pt}
\tablehead{
\colhead{Name}& 		
\colhead{Type}& 	
\colhead{V$_{\rm mag}$}&	
\colhead{Slit}&
\colhead{Binning}&
\colhead{Data Sets}&				
\colhead{Exp. (s)}&	
\colhead{Fringe Data Set}&	
\colhead{PID}&	
\colhead{Date-Obs}\\
}
\startdata
HD141680 & G8 III	& 5.23 & 0\farcs1 & x1 & O67117020			& 2.2 & O67117030	& 8591 & 2001-04-04 \\
		  &		&	   & 0\farcs2 & x1 & O67117040			& 2.2	 & O67117050	&	     &			   \\
HR6770	  & G8 III	& 4.65 & 0\farcs1 & x2 & O4AN18010,20,30,40,50	& 2.2	 & O4AN18060	& 7388 & 1998-09-06 \\
		  &		&	   & 0\farcs2 & x2 & O4AN18070,80,90,A0,B0	& 2.2	 & O4AN180C0	&	     &			   \\
HR7576	  & K3 III	& 5.03 & 0\farcs1 & x2 & O4AN17010,20,30,40,50 	& 3.2	 & O4AN17060	& 7388 & 1998-05-24 \\
		  &		&	   & 0\farcs2 & x2 & O4AN17070,80,90,A0,B0 	& 3.2	 & O4AN170C0	&	     &			   \\
\enddata
\tablecomments{Suitable LOSVD late-type template standard stars available in the {\it HST} archive that include contemporaneous flat fields necessary for defringing. 
`Slit' is the width of the STIS long-slit used in arc-seconds. `Binning' is the on-chip binning used in the dispersion direction. `PID' is the {\it HST} program during which 
these observations were taken.}
\end{deluxetable*}

\begin{deluxetable*}{lcccccccc}
\vspace{-0.9cm}
\tablewidth{510pt}
\tabletypesize{\footnotesize}
\tablecaption{Available STIS CaII Stellar Absorption Line Data for non-Barred Galactic Nuclei\label{tab:sample}}
\tablewidth{0pt}
\tablehead{
\colhead{Name}&
\colhead{Class}&
\colhead{D (ref.)}&
\colhead{$\log M_\bullet$ (ref.)}&
\colhead{Slit}&
\colhead{Binning}&
\colhead{Exp. Time}&
\colhead{Prop. ID}&
\colhead{$\sigma_\ast$}\\
\colhead{}&
\colhead{}&
\colhead{(Mpc)}&
\colhead{($M_\odot$)}&
\colhead{}&
\colhead{}&
\colhead{(s)}&
\colhead{}&
\colhead{(\kms)}
}
\startdata
NGC~205 	& E5 pec 		& 0.82 (1) 		& $<4.34$ (s, 7) 				& $0\farcs1$ & x1 & 24132 	& 9448 & $31\pm5$		\\
NGC~221 	& cE2 		& 0.81 (2) 		& $6.40_{-0.10}^{+0.08}$ (s, 8)		& $0\farcs1$ & x1 & 4899 		& 7566 & $72\pm2$ 		\\
NGC~224 	& SA(s)b 		& 0.79 (1) 		& $8.15_{-0.11}^{+0.21}$ (s, 9)		& $0\farcs1$ & x1 & 20790 	& 8018 & $170\pm5$ 	\\
NGC~598 	& SA(s)cd 	& 0.81 (1) 		& $<$3.48 (s, 10) 				& $0\farcs1$ & x1 & 7380 		& 8018 & $37\pm16$ 	\\
NGC~821 	& E6 		& 24.10 (2	) 	& $7.93_{-0.23}^{+0.15}$ (s, 11)	& $0\farcs1$ & x2 & 27478 	& 7388 & $200\pm3$  	\\
NGC~1374 	& E3 		& 19.77 (2) 	& \nodata		   	     			& $0\farcs1$ & x1 & 7372 		& 9107 & $186\pm4$   	\\
NGC~1700 	& E4 		& 44.26 (2) 	& \nodata		  	     			& $0\farcs2$ & x2 & 7170 		& 7566 & $239\pm4$   	\\
NGC~2434 	& E0-1 		& 21.58 (2) 	& \nodata		   	     			& $0\farcs2$ & x1 & 28400 	& 9107 & $187\pm7$  	\\
NGC~2778 	& E 			& 22.91 (2) 	& $7.15_{-0.45}^{+0.19}$ (s, 12)  	& $0\farcs1$ & x2 & 15026 	& 7388 & $162\pm3$	\\
NGC~2784 	& SA(s) 		& 9.82 (2) 		& \nodata		   	     			& $0\farcs1$ & x1 & 8740 		& 8591 & $225\pm7$   	\\
NGC~2841 	& SA(s)b 		& 14.06 (3) 	& \nodata		  	     			& $0\farcs2$ & x2 & 7990 		& 8022 & $206\pm4$  	\\
NGC~3031 	& SA(s)ab 	& 3.91 (2) 		& $7.88_{-0.07}^{+0.11}$ (g, 13)  	& $0\farcs2$ & x1 & 39200 	& 7350 & $162\pm3$  	\\
NGC~3115 	& S0- 		& 9.68 (2) 		& $8.96_{-0.16}^{+0.33}$ (s, 14)  	& $0\farcs1$ & x1 & 7360 		& 7566 & $267\pm4$   	\\
NGC~3585* 	& E7/S0 		& 20.04 (2) 	& $8.53_{-0.08}^{+0.16}$ (s, 23) 	& $0\farcs1$ & x1 & 12241 	& 9107 & $206\pm7$   	\\
NGC~3593 	& SA(s) 		& 9.72 (4) 		& \nodata		   	     			& $0\farcs1$ & x1 & 5990 		& 8591 & $54\pm7$   	\\
NGC~3607 	& SA(s) 		& 22.80 (2) 	& $8.08_{-0.18}^{+0.12}$ (s, 23) 	& $0\farcs2$ & x1 & 26616 	& 9107 & $224\pm10$  	\\
NGC~3608 	& E2 		& 22.91 (2) 	& $8.28_{-0.17}^{+0.18}$ (s, 12)  	& $0\farcs2$ & x2 & 12950 	& 7388 & $192\pm4$   	\\
NGC~3706* 	& SA(rs) 		& 38.02 (5) 	& \nodata		   	     			& $0\farcs1$ & x1 & 15174 	& 8687 & $270\pm8$  	\\
NGC~3998 	& SA(r) 		& 14.13 (2) 	& $8.91_{-0.12}^{+0.09}$ (g, 15)  	& $0\farcs2$ & x1 & 21840 	& 7350 & $280\pm15$ 	\\
NGC~4026 	& S0 		& 13.61 (2) 	& $8.32_{-0.13}^{+0.09}$ (s, 23)	& $0\farcs1$ & x1 & 9973 		& 9107 & $178\pm4$  	\\
NGC~4278 	& E1-2 		& 16.07 (2) 	& \nodata		  	     			& $0\farcs2$ & x2 & 17502 	& 7350 & $237\pm5$ 	\\
NGC~4291 	& E3 		& 26.18 (2) 	& $8.49_{-0.59}^{+0.10}$ (s, 12)  	& $0\farcs2$ & x2 & 16013 	& 7388 & $285\pm6$  	\\
NGC~4382 	& SA(s)0+ pec 	& 17.86 (6) 	& \nodata		  	     			& $0\farcs2$ & x1 & 18794 	& 9107 & $179\pm5$ 	\\
NGC~4473 	& E5 		& 15.70 (2) 	& $8.04_{-0.56}^{+0.14}$ (s, 12)  	& $0\farcs2$ & x2 & 12840 	& 7388 & $179\pm3$  	\\
NGC~4486 	& E+0-1 pec 	& 17.22 (6) 	& $9.81_{-0.04}^{+0.03}$ (g, 16)  	& $0\farcs2$ & x2 & 25300 	& 7567 & $334\pm5$  	\\
NGC~4486A 	& E2 		& 18.28 (6) 	& $7.11_{-0.41}^{+0.21}$ (s, 17)  	& $0\farcs2$ & x1 & 21112 	& 8687 & $135\pm4$	\\
NGC~4486B 	& cE0 		& 16.29 (6) 	& $8.78_{-0.18}^{+0.17}$ (s, 18)  	& $0\farcs2$ & x2 & 14530 	& 7566 & $169\pm4$	\\
NGC~4552 	& E 			& 15.85 (6) 	& $8.68_{-0.08}^{+0.07}$ (s, 19)	& $0\farcs2$ & x2 & 7490 		& 8022 & $253\pm3$  	\\
NGC~4564 	& E6 		& 15.85 (6) 	& $7.75_{-0.07}^{+0.02}$ (s, 12)  	& $0\farcs1$ & x2 & 15060 	& 7388 & $157\pm3$ 	\\
NGC~4621 	& E5 		& 14.93 (6) 	& $8.60_{-0.07}^{+0.06}$ (s, 19)  	& $0\farcs2$ & x1 & 4340		& 8018 & $225\pm3$ 	\\
NGC~4649 	& E2 		& 17.30 (2) 	& $9.30_{-0.15}^{+0.08}$ (s, 12) 	& $0\farcs2$ & x2 & 46467 	& 7388 & $335\pm5$ 	\\
NGC~4697 	& E6 		& 11.75 (2) 	& $8.23_{-0.03}^{+0.05}$ (s, 12)  	& $0\farcs1$ & x2 & 26920	& 7388 & $171\pm2$ 	\\
NGC~4736 	& (R)SA(r)ab 	& 5.20 (2) 		& \nodata		  	     			& $0\farcs1$ & x1 & 5769 		& 8591 & $104\pm4$ 	\\
NGC~4742 	& E4 		& 15.49 (2) 	& $7.15_{-0.20}^{+0.11}$ (s, 20)  	& $0\farcs2$ & x1 & 7130 		& 8018 & $108\pm4$  	\\
NGC~4826 	& (R)SA(rs)ab 	& 7.48 (2) 		& \nodata		  	     			& $0\farcs1$ & x1 & 5990	 	& 8591 & $92\pm4$   	\\
NGC~5033 	& SA(s)c 		& 17.22 (5) 	& \nodata		  	     			& $0\farcs2$ & x1 & 10300 	& 9776 & $131\pm7$   	\\
NGC~5055 	& SA(rs)bc 	& 8.75 (5) 		& $8.93_{-0.11}^{+0.09}$ (g, 21)  	& $0\farcs1$ & x1 & 4190 		& 8591 & $101\pm3$ 	\\
NGC~5102 	& SA0- 		& 4.00 (2) 		& \nodata		  	     			& $0\farcs1$ & x1 & 4592 		& 8591 & \nodata 		\\
NGC~5576 	& E3 		& 25.47 (2) 	& $8.26_{-0.11}^{+0.06}$ (s, 23)	& $0\farcs1$ & x1 & 7138 		& 9107 & $171\pm5$  	\\
NGC~7213 	& SA(s) 		& 22.70 (5) 	& \nodata		  	     			& $0\farcs1$ & x1 & 7603 		& 9107 & $163\pm9$   	\\
NGC~7332 	& S0 pec sp 	& 24.89 (2) 	& $7.11_{-0.21}^{+0.19}$ (s, 22)  	& $0\farcs2$ & x1 & 9870 		& 7566 & $125\pm3$   	\\
NGC~7457 	& SA(rs)0-? 	& 13.24 (2) 	& $6.54_{-0.22}^{+0.12}$ (s, 12)  	& $0\farcs1$ & x2 & 15282 	& 7388 & $69\pm4$		\\
\enddata
\tablecomments{Non-barred galactic bulges with CaII stellar dynamics observed by STIS (G750M, 8561\AA). Galaxies marked with * also have 0\farcs2 slit widths available. ``Class'': NED classifications. Method codes: s = stellar dynamics; g = gas dynamics. The values of $\sigma_\ast$ are those listed as central velocity dispersions by Hyperleda (http://leda.univ-lyon1.fr/; \citealt{2003A&A...412...45P}). ``Slit" is the width of the STIS slit used. ``Binning" is the on chip binning used in the dispersion direction. ``Exp.Time" is the total exposure time.\\
{\bf References:} 
(1) \cite{2005MNRAS.356..979M}; 
(2) \cite{2001ApJ...546..681T}; 
(3) \cite{2001ApJ...559..243M}; 
(4) \cite{1992A&A...257..437W}; 
(5) NED (Virgo+GA+Shapley corrected Hubble flow distances); 
(6) \cite{2007ApJ...655..144M}; 
(7) \cite{2005ApJ...628..137V};
(8) \cite{2001ApJ...550..668J};
(9) \cite{2005ApJ...631..280B};
(10) \cite{2001Sci...293.1116M};
(11) \cite{2004astro.ph..3257R};
(12) \cite{2003ApJ...583...92G};
(13) \cite{2003AJ....125.1226D};
(14) \cite{1999MNRAS.303..495E};
(15) \cite{2012ApJ...753...79W};
(16) \cite{2009ApJ...700.1690G};
(17) \cite{2007MNRAS.379..909N};
(18) \cite{1997ApJ...482L.139K};
(19) \cite{2008MNRAS.386.2242H};
(20) \cite{2002ApJ...574..740T};
(21) \cite{2004A&A...420..147B};
(22) \cite{2004ApJ...604L..89H};
(23) \cite{2009ApJ...695.1577G}.}
\end{deluxetable*}

There are predominantly two methods for directly estimating \m; gas kinematics and stellar dynamics. While not all galaxies contain nuclear gas disks \citep{2001AJ....121.2928T}, the Keplerian kinematics of gas is relatively easy to model (e.g., \citealt{1996ApJ...470..444F,2001ApJ...549..915M}) provided the gas is dominated by the gravitational potential of the SBH and not subject to significant inflow, outflow or turbulence. In contrast, stellar dynamics (e.g., \citealt{1998ApJ...493..613V,1999ApJS..124..383C,2002MNRAS.335..517V,2003ApJ...583...92G,2004ApJ...602...66V,2010MNRAS.401.1770V}), do not suffer from non-gravitational influences and can theoretically be applied to all galaxies. However, unless the root-mean-square (rms) stellar velocities show a clear rise near the center, which is only true for a handful of galaxies, stellar dynamical models require significantly more data in order to constrain \m\ and (compared with emission-line data) more observing time must be invested to accurately determine the line-of-sight velocity distributions (LOSVDs), particularly in the low-surface-brightness cores of bright galaxies.

A fundamental step in obtaining more high-accuracy direct estimates of \m\ is to compile a list of galaxies that have the greatest potential for success. The \cite{2003AJ....126..742H} Space Telescope Imaging Spectrograph (STIS) atlas of nearby spiral galaxies has enabled the identification of nuclear disks ideal for further  \m\ estimates using gas kinematics, however there does not exist a similar atlas for stellar dynamics. Consequently, we present such an atlas here. For each of the galaxies in the chosen sample we provide the signal-to-noise (S/N) profile across the STIS slit. In addition, we present the inferred moments of the LOSVDs derived from the highest S/N spectra in each galaxy. Finally, we identify those galaxies in the {\it HST} archive that may produce further estimates of \m\ from stellar dynamics. In \S~\ref{sample} we describe the sample. The details of the data reduction are in \S~\ref{data}. The results are compiled in \S~\ref{results} and discussed in \S~\ref{disc}. \S~\ref{cons} concludes. 

\section{Sample Selection}\label{sample}

In order to determine the relative suitability of the available data for SBH modeling we use two approaches. First, we calculate the S/N profiles of each galaxy across the STIS slit. Second, we match appropriate stellar templates to the CaII absorption features and determine how well the central LOSVDs can be recovered. Therefore, standard stars that have been collected using identical set ups to the galaxies observed must also be recovered from the {\it HST} archive. Late-type G, K and M stars are ideal to best match the stellar populations at the centers of galactic bulges. Currently, there are eight stellar templates in the archive that have been observed with the same set up as the galaxy sample. However, as detailed in \S~\ref{data}, only 3 of these have the appropriate contemporaneous calibration flat fields necessary to satisfactorily remove near-infrared fringing. The details of these standard stars are presented in Table~\ref{tab:standards}.

We have selected all NASA Extra-galactic Database (NED)\footnote{http://nedwww.ipac.caltech.edu/} identified non-barred galactic bulges in the {\it HST} archive that have G750M STIS long-slit spectra ($52''\times0\farcs1$ or $52''\times0\farcs2$) centered at 8561\AA. The high spatial resolution and sensitivity of {\it HST} make it the natural choice from which to compile a sample of galactic bulges suitable for stellar dynamical modeling of the central SBH. Indeed, most \m\ estimates to date are a result of {\it HST} observations, many of which are included in this sample for comparison purposes. We exclude barred-spirals as dynamical models that can accommodate triaxial potentials are still under development. To avoid any possible selection bias \citep{2010ApJ...711L.108B}, we do not discard any galaxy based on spatially resolving the predicted SBH sphere of influence. The G750M grating, centered at 8561\AA, ensures the CaII stellar absorption triplet (8500\AA, 8544\AA\ \& 8665\AA\ vacuum) can be observed at the highest possible spectral resolution. This enables a better determination of the LOSVDs needed for dynamical models. The archive does contain 3 additional galaxies that have been observed at the wavelengths of the Mgb triplet (G430M), however the CaII triplet is favored due to a reduced impact from standard star template mismatching. This selection results in 42 galaxies, the details of which are compiled in Table~\ref{tab:sample}.

\section{Data Reduction}\label{data}

After being passed through the latest CALSTIS pipeline for on-the-fly-recalibration using the best reference files, all data were retrieved from the {\it HST} archive. At wavelengths longer than $\sim$7000\AA, STIS experiences significant fringing from multiple reflections between the two surfaces of its CCD. More extensive details of the issue can be found in several Instrument Science Reports, i.e., ISR 97-16 \citep{1997stis.rept...16W}, ISR 98-19 \citep{1998stis.rept...19G}, ISR 98-29 \citep{1998stis.rept...29G}. For the G750M grating centered at 8561\AA\ the amplitude of the fringes can be between 10-15\% for observations with a S/N $>$ 50; the relative effects of fringing are much more difficult to detect in low S/N data. Thus, fringing will have an impact on the standard star observations, and potentially the galaxy data. In addition, due to the non-repeatability in the position of some of the STIS mechanisms, the fringe patterns are not stationary with time, i.e., a standard fringe flat field cannot be used. Therefore, in every case, the appropriate contemporaneous fringe flats were also retrieved from the {\it HST} archive; these are not normally included in the calibration files distributed with archived data. Care must be taken to ensure the most appropriate fringe flat field is used. For example, there are differences in the removal of the fringe pattern when observing point and diffuse sources, and when using different slit widths. This will be more thoroughly detailed in the following sections.  

\begin{figure}
\includegraphics*[width=245pt]{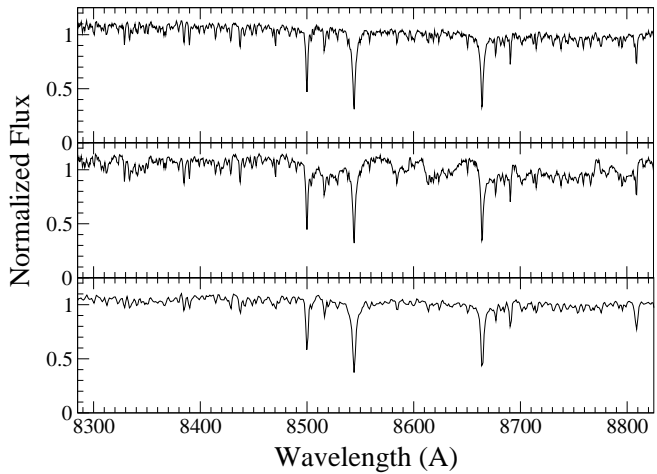}
\caption{[Top] HD141680 un-binned stellar template through the 0\farcs2 slit averaged across 7 rows. [Middle] The same HD141680 spectrum without de-fringing.
[Bottom] Composite x2 binned stellar template through the 0\farcs2 slit averaged across 9 rows.}
\label{fig:standards}
\end{figure}

\subsection{Standard Stars}

All three standards were observed using the 0\farcs1 and 0\farcs2 slit widths. The data from the two slit widths are treated separately so that an appropriate template can be generated to match the slit widths used on the galaxies. Contemporaneous fringe flats were recovered that are optimized for these slits widths and for point sources, i.e., fringes observed through the $0\farcs2\times0\farcs06$ slit for the $52''\times0\farcs1$ observations, and $0\farcs3\times0\farcs09$ for the $52''\times0\farcs2$ observations (the use of these short-slit fringe flats better mimic the point spread function than the long-slit flats, \citealt{1998stis.rept...19G}). In addition, the data for HD141680 was not binned on-chip, whereas the HR6770 and HR7576 data was binned by 2 pixels in the dispersion direction. The galaxy sample contains both un-binned and 2 pixel binned data, so the binned and un-binned template stars are also treated separately. 

After the pipeline reduced data had been defringed, large amplitude variations were still noted in the central row of the spectra. This is attributed to a small residual misalignment between the dispersion direction of the spectra and the pixel rows. This was confirmed by performing a column-by-column trace of the spectrum peak. The peak positions correlate with the amplitude variation in the central row. This residual mis-alignment can be removed by re-centering the spectra column-by-column to a common position, or by collapsing the data across the slit width (7 and 9 pixels for the 0\farcs1 and 0\farcs2 slits, respectively). There is no significant difference between the resulting spectra produced either way, so the intra-slit collapse was used for the sake of simplicity. The improvement to the template spectra as a result of defringing is demonstrated in Figure~\ref{fig:standards} for HD141680. The standard deviation of the residual spectra between the defringed and final spectra is 5\%. 

HD141680 was only observed at a single spatial slit position. However, in order to map the line profiles across the slit, spatial steps across the positions of HR6770 and HR7576 have been performed, thus producing five spectra at different locations for each of these standards. In each visit to these standards, the 0\farcs1 slit data was gathered first. Therefore, in both cases the brightest spectra come from the slit with zero spatial offset ({\tt POSTARG1 = 0.00}). However, the brightest spectra are found in the {\tt POSTARG1 = 0.04} position for the 0\farcs2 slits. In both cases it was found that the guide star acquisition had failed, and that the pointing of {\it HST} had drifted slightly over the $\sim16$ minute period between the peak-up and beginning of the 0\farcs2 slit observations. Only data from the brightest spectra are used to generate the stellar template.  

For the galaxy data that is un-binned, to preserve the maximum spectral resolution there is only one choice of template star to estimate the LOSVDs (HD141680). However, for the galaxy data binned by 2 pixels there is HR6770 and HR7576. Plus, the HD141680 data can be additionally binned by 2 dispersion pixels to create a third useful standard. To reduce the possibility of template mis-matching, a composite template is produced by normalizing each individual standard spectra and averaging them together. Thus, we are left with 4 templates. Templates A and B are the un-binned HD141680 data using the 0\farcs1 and 0\farcs2 data, respectively. Templates C and D are the 2 pixel binned data generated from a composite of all three standards using the 0\farcs1 and 0\farcs2 data, respectively. Two templates (B and D) are presented in Figure~\ref{fig:standards} for the 0\farcs2 slit. Gaussian fits to the prominent CaII triplet shows the centers of the lines to be within 10~\kms\ of the vacuum values expected, indicative to the level of precision in the data reduction techniques and the heliocentric velocity corrections."

\subsection{Galaxies}

For extended sources, the appropriate contemporaneous fringe flat should be taken through the same slit configuration. Thus, there is a minor difference in the defringing process for galaxies. However, only a marginal improvement in flat fielding is found for G750M data with S/N $<$ 50 \citep{1997stis.rept...16W}, and the peak S/N found in this sample is 26.7 (NGC 3031). The defringing process requires the fringe patterns in the data and fringe flat field to be both spatially aligned and matched in amplitude. Therefore, an RMS minimization is performed to shift and scale the fringe flats to the data. For the galaxy spectra, the defringing process produced multiple local minima in the RMS values calculated for a wide range of non-physical spatial shifts and amplitudes. Consequently, we find that defringing low S/N G750M data with contemporaneous flat fields produces no improvement in these data. 

In many cases, the galaxy spectra have been gathered using steps (spacecraft pointing offsets) across and along the slit direction. As with the standards, we do not present the spatially offset spectra (but do note when such data is available) and simply show the spectra with the highest S/N. However, spectra that have been stepped along the slit direction can be co-aligned and combined. In these cases the position of the spectrum on the detector was traced by a column-by-column 1-D fit to the position of the continuum. The relative offsets between each spectra were then calculated and used to shift each spectrum back to a common position. The co-aligned spectra were then median combined to remove any residual post CR-split cosmic rays and hot pixels. 

\section{Signal-to-Noise Ratio}\label{ston}

The signal-to-noise ratio in a spectrum is one obvious, and frequently used, method for assigning a measure of quality to data. Therefore, in determining the suitability of galaxies for future dynamical modeling aimed at constraining \m, calculating the S/N in the STIS spectra is a natural first step. However, this seemingly trivial exercise is not necessarily as straightforward as one may think; one cannot simply take the square-root of the detector data number as this would neglect, for example, background counts and read noise. None of the standard post-pipeline data reduction tasks specifically deal with data quality or error propagation. One could consider calculating the ratio of the mean flux and standard deviation along the spectra. However, such calculations would be contaminated by absorption and emission lines at varying wavelengths and widths.

\begin{figure}
\plotone{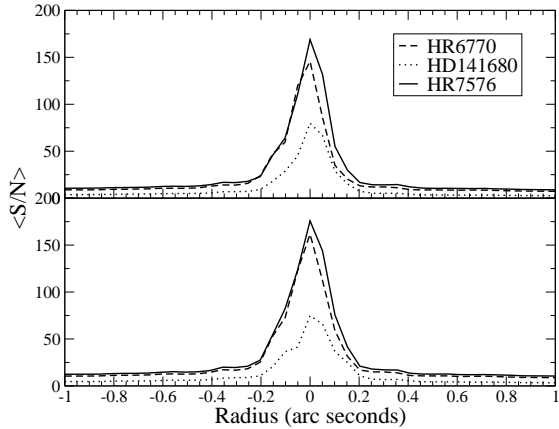}
\caption{Mean S/N ($<$S/N$>$) profiles of the three standard stars through the 0\farcs1 slit [top] and 0\farcs2 [bottom]. The spatial sampling is 0\farcs05078.}
\label{fig:snrstars}
\end{figure}

\begin{deluxetable}{lcccc}
\tablewidth{245pt}
\tabletypesize{\footnotesize}
\tablecaption{S/N Tests for the Standard Stars\label{tab:snrcheck}}
\tablewidth{0pt}
\tablehead{
\colhead{Name}& 
\colhead{Slit}&		
\colhead{SNRMAX}& 	
\colhead{$\rm S/N_{peak}$}&	
\colhead{$<$S/N$>$}\\
}
\startdata
HD141680 & 0\farcs1	& 93.8 	& 95.6 	& 80$\pm$7	\\
		  & 0\farcs2	& 96.4 	& 98.2 	& 75$\pm$11	\\
HR6770	  & 0\farcs1	& 166.7 	& 169.1 	& 146$\pm$9	\\
		  & 0\farcs2	& 173.2 	& 185.1 	& 162$\pm$10	\\
HR7576	  & 0\farcs1	& 171.8	& 192.8 	& 169$\pm$10	\\
		  & 0\farcs2	& 170.5	& 203.3 	& 176$\pm$10	\\
\enddata
\tablecomments{Signal-to-noise ratios for the standard stars. SNRMAX is the pipeline header parameter for the reduced templates. It is the maximum S/N calculated in a single pixel. $\rm S/N_{peak}$ is the maximum per-pixel S/N calculated from our S/N arrays. $<$S/N$>$ is the mean and standard deviation of the S/N along the peak spectra from our S/N array. }
\end{deluxetable}

However, to include both the science and reference file errors, the pipeline does propagate through the per pixel statistical errors estimated from the bias subtracted observed data number (counts), the gain and the read noise. These data are assigned to an error array that is included in the second extension of the pipeline provided data files. So, it is possible to assign a per pixel S/N based on the ratio of the science array to the error array. Indeed, this is how the pipeline assigns the header parameters that show the maximum, minimum and mean S/N ($<$S/N$>$) in a processed data array. Unfortunately, this does not provide the S/N in an individual spectra. Therefore, the error arrays have been extracted, squared, aligned using the same offsets calculated for the science arrays, median combined and then square rooted. The S/N arrays are then created simply by dividing the co-aligned and combined science array by the equivalent error array. This provides an array from which we can extract the S/N in any pixel, or along any spectrum. 

Table~\ref{tab:snrcheck} presents the S/N for the standard stars. The maximum per pixel values from the S/N arrays are consistent with the maximum per pixel S/N calculated by the pipeline. The small differences will be due to the defringing process that is not accounted for in the pipeline calculated S/N. However, these S/N values are based on single pixels so, while they do demonstrate that our S/N array is consistent with the pipeline S/N calculations, they do not provide any information on the S/N present in the spectra. Consequently, the mean and standard deviation of the S/N from the brightest template spectra are also presented in Table~\ref{tab:snrcheck}. These values are lower than the peak per pixel S/N by $\lesssim$15\%, which can be attributed to including the lower S/N across the absorption lines. In addition, the $<$S/N$>$ ratios calculated here are consistent with those expected from the STIS exposure time calculator\footnote{http://etc.stsci.edu/etc/input/stis/spectroscopic/} (ETC). The spatial $<$S/N$>$ profiles for each standard star are presented in Figure~\ref{fig:snrstars}. 

\begin{figure}
\plotone{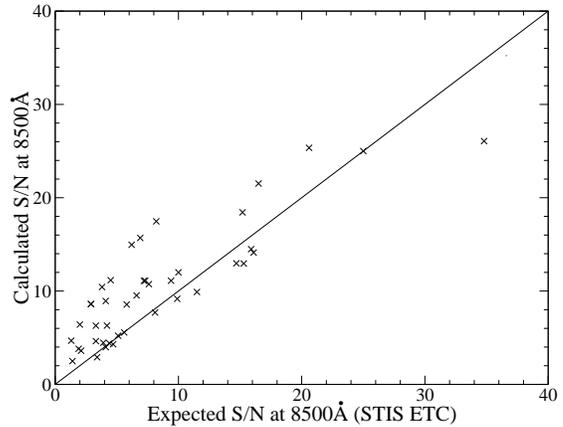}
\caption{Comparison between the galaxy S/N at 8500\AA~ calculated here and the S/N expected from the STIS ETC at 8500\AA. The solid line marks the 1:1 relation.}
\label{fig:snretc}
\end{figure}

In column 2, Table~\ref{tab:results} lists the $<$S/N$>$ along the peak galaxy spectra. An interesting check is to compare our calculated S/N values to those predicted by the STIS ETC. Figure~\ref{fig:snretc} presents such a comparison. To be consistent with the S/N arrays, the ETC S/N values were estimated using the peak galaxy flux in a single pixel at 8500\AA, and the median of the S/N values that resulted from the individual (pre-combined) exposure times. Figure~\ref{fig:snretc} shows good agreement between the two methods. Some scatter is expected because the ETC S/N calculations have used updated sensitivity values as the detector has aged. As LOSVDs over a spatially extended region are necessary to constrain \m\ models the S/N spatial profiles are also presented for all galaxies in the Appendix.

\begin{deluxetable*}{lcccccccc}
\vspace{-0.8cm}
\tabletypesize{\footnotesize}
\tablecaption{Results from the MPL Routine\label{tab:results}}
\tablewidth{510pt}
\tablehead{
\colhead{Name}&
\colhead{Template}&
\colhead{$<$S/N$>$} &
\colhead{ISE}&
\colhead{$\alpha$}&
\colhead{$v_c$}&
\colhead{$\sigma_c$}&
\colhead{$h_3$}&
\colhead{$h_4$} \\
\colhead{1}&
\colhead{2}&
\colhead{3}&
\colhead{4}&
\colhead{5}&
\colhead{6}&
\colhead{7}&
\colhead{8}&
\colhead{9}
}
\startdata
NGC~205*  & A & 4.5    &  3.82    & 3.4     & $-232^{+3}_{-4}$    & $21^{+5 }_{-3}$    & $0.03^{+0.03}_{-0.09}$  & $0.05^{+0.03}_{-0.04}$ \\
NGC~221*  & A & 25.2   &  1.1	  & 5.4     & $-195^{+19}_{-6}$   & $165^{+21}_{-29}$  & $0.03^{+0.06}_{-0.06}$  & $0.1^{+0.05}_{-0.06}$ \\
NGC~224*  & A & 8.6    &  1.64    & 6.7     & $-175^{+27}_{-14}$  & $168^{+25}_{-22}$  & $-0.15^{+0.08}_{-0.04}$ & $0.03^{+0.06}_{-0.04}$ \\
NGC~598*  & A & 13.7   &  1.62    & 2.9     & $-180^{+2}_{-2}$    & $19^{+3 }_{-2}$    & $0.07^{+0.05}_{-0.08}$  & $0.02^{+0.04}_{-0.02}$ \\
NGC~821*  & B & 8.3    &  1.81    & 8.9     & $1734^{+27}_{-70}$  & $354^{+35}_{-58}$  & $0.11^{+0.05}_{-0.11}$  & $0.03^{+0.06}_{-0.06}$ \\
NGC~1374  & A & 3.7    &  6.84    & 9	    & $1244^{+122}_{-44}$ & $338^{+31}_{-60}$  & $-0.09^{+0.16}_{-0.08}$ & $-0.04^{+0.08}_{-0.07}$ \\
NGC~1700  & D & 14.1   &  1.42    & 10.3    & $3732^{+82}_{-36}$  & $468^{+5 }_{-100}$ & $-0.01^{+0.06}_{-0.05}$ & $0.01^{+0}_{-0.05}$ \\
NGC~2434  & C & 6.2    &  2.6	  & 10      & $1496^{+44}_{-25}$  & $369^{+7 }_{-18}$  & $-0.08^{+0.05}_{-0.02}$ & $-0.06^{+0.06}_{0}$ \\
NGC~2778* & B & 6.2    &  2.97    & 8.1     & $2058^{+33}_{-45}$  & $191^{+52}_{-20}$  & $0^{+0.1}_{-0.11}$	 & $-0.06^{+0.1}_{-0.03}$ \\
NGC~2784  & A & 5.1    &  3.69    & 8.2     & $617 ^{+42}_{-58}$  & $265^{+62}_{-48}$  & $0^{+0.07}_{-0.08}$	 & $0.09^{+0.05}_{-0.08}$ \\
NGC~2841  & D & 13     &  1.51    & 9.6     & $692 ^{+26}_{-55}$  & $327^{+21}_{-52}$  & $0.12^{+0}_{-0.11}$	 & $0.06^{+0.02}_{-0.07}$ \\
NGC~3031  & C & 26.7   &  \nodata & \nodata & \nodata  	  & \nodata	       & \nodata		 & \nodata		  \\
NGC~3115  & A & 13.3   &  1.52    & 11      & $674 ^{+43}_{-66}$  & $466^{+51}_{-17}$  & $0^{+0.03}_{-0.03}$	 & $-0.02^{+0.02}_{-0}$ \\
NGC~3585* & A & 5.5    &  3.38    & 8.5     & $1434^{+74}_{-46}$  & $312^{+36}_{-97}$  & $-0.15^{+0.12}_{-0.02}$ & $0.05^{+0.05}_{-0.1}$  \\
NGC~3593  & A & 4.3    &  11.12   & 6.5     & $636 ^{+7}_{-22}$   & $72^{+15}_{-12}$   & $0.11^{+0}_{-0.18}$	 & $0.01^{+0.03}_{-0.03}$ \\
NGC~3607* & C & 2.9    &  3.82    & 9.6     & $899 ^{+30}_{-28}$  & $253^{+4 }_{-23}$  & $0.02^{+0.01}_{-0.05}$  & $-0.03^{+0.03}_{-0.01}$ \\
NGC~3608* & D & 9.1    &  1.45    & 11.4    & $1170^{+38}_{-35}$  & $338^{+25}_{-56}$  & $0^{+0}_{-0}$  	 & $0^{+0}_{-0.01}$ \\
NGC~3706  & A & 3.5    &  4.65    & 7.8     & $3002^{+22}_{-49}$  & $173^{+29}_{-21}$  & $0.05^{+0}_{-0.12}$	 & $-0.04^{+0.08}_{-0.03}$ \\
NGC~3998  & C & 19.9   &  \nodata & \nodata & \nodata  	  & \nodata	       & \nodata		 & \nodata		 \\
NGC~4026* & A & 8.4    &  2.36    & 9.6     & $974 ^{+21}_{-42}$  & $352^{+26}_{-35}$  & $0.01^{+0.02}_{-0.04}$  & $-0.01^{+0.01}_{-0.03}$ \\
NGC~4278  & D & 5.6    &  \nodata & \nodata & \nodata  	  & \nodata	       & \nodata		 & \nodata		 \\
NGC~4291* & D & 8.3    &  1.49    & 12      & $1892^{+56}_{-91}$  & $515^{+21}_{-67}$  & $-0.01^{+0.01}_{-0.01}$ & $-0.01^{+0.01}_{-0}$ \\
NGC~4382  & C & 4.1    &  3.08    & 11.6    & $693 ^{+6}_{-16}$   & $180^{+1}_{-1}$   & $0^{+0}_{-0}$  	 & $0^{+0}_{-0}$ \\
NGC~4473* & D & 8.8    &  1.95    & 8.5     & $2305^{+37}_{-40}$  & $253^{+17}_{-47}$  & $0.05^{+0.08}_{-0.09}$  & $-0.06^{+0.13}_{-0}$ \\
NGC~4486  & D & 16.8   &  \nodata & \nodata & \nodata  	  & \nodata	       & \nodata		 & \nodata		 \\
NGC~4486A & C & 3.4    &  3.04    & 7.3     & $798 ^{+25}_{-15}$  & $153^{+22}_{-20}$  & $0.04^{+0.07}_{-0.09}$  & $0.07^{+0.04}_{-0.06}$ \\
NGC~4486B & D & 10.8   &  4.37    & 8.6     & $1542^{+47}_{-99}$  & $254^{+66}_{-40}$  & $0.04^{+0.11}_{-0.15}$  & $-0.04^{+0.11}_{-0.05}$ \\
NGC~4552  & D & 9.5    &  2.32    & 12      & $356 ^{+39}_{-50}$  & $259^{+1 }_{-9}$   & $0^{+0}_{-0}$  	 & $0^{+0}_{-0}$ \\
NGC~4564* & B & 10.5   &  2.94    & 11.5    & $1113^{+18}_{-22}$  & $159^{+1}_{-2}$   & $0^{+0}_{-0}$  	 & $0^{+0}_{-0}$ \\
NGC~4621  & C & 13.9   &  \nodata & \nodata & \nodata  	  & \nodata	       & \nodata		 & \nodata		 \\
NGC~4649* & D & 4.3    &  2.12    & 10.4    & $822 ^{+136}_{-54}$ & $551^{+35}_{-82}$  & $-0.1^{+0.1 }_{-0.03}$  & $0.01^{+0.03}_{-0.05}$ \\
NGC~4697* & B & 9.6    &  1.34    & 8.4     & $1290^{+17}_{-43}$  & $224^{+23}_{-24}$  & $0.1^{+0.03}_{-0.09}$   & $0.01^{+0.05}_{-0.04}$ \\
NGC~4736  & A & 9.7    &  3.42    & 6.7     & $343 ^{+15}_{-13}$  & $101^{+12}_{-14}$  & $-0.06^{+0.07}_{-0.05}$ & $0.05^{+0.02}_{-0.05}$ \\
NGC~4742  & C & 19.8   &  1.32    & 5.8     & $1317^{+11}_{-16}$  & $107^{+14}_{-11}$  & $0.04^{+0.06}_{-0.08}$  & $-0.03^{+0.06}_{-0.05}$ \\
NGC~4826  & A & 4.1    &  5.92    & 6.4     & $442 ^{+17}_{-121}$ & $112^{+5 }_{-35}$  & $0.03^{+0.09}_{-0.12}$  & $0.09^{+0.01}_{-0.11}$ \\
NGC~5033  & C & 10     &  2.49    & 10.6    & $890 ^{+23}_{-19}$  & $242^{+2 }_{-24}$  & $-0.01^{+0.01}_{-0}$	 & $0^{+0}_{-0.01}$ \\
NGC~5055  & A & 12.4   &  2.39    & 5.8     & $493 ^{+22}_{-13}$  & $119^{+18}_{-22}$  & $-0.15^{+0.1}_{-0.06}$  & $0.02^{+0.08}_{-0.07}$ \\
NGC~5102  & A & 24.4   &  1.63    & 5.6     & $496 ^{+8}_{-13}$   & $97^{+6 }_{-13}$   & $-0.06^{+0.17}_{-0.05}$ & $-0.06^{+0.17}_{-0}$ \\
NGC~5576* & A & 4      &  6.83    & 8.8     & $1607^{+31}_{-72}$  & $238^{+39}_{-41}$  & $0.14^{+0.01}_{-0.15}$  & $0.03^{+0.04}_{-0.07}$ \\
NGC~7213  & A & 11.4   &  \nodata & \nodata & \nodata  	  & \nodata	       & \nodata		 & \nodata		 \\
NGC~7332  & C & 14     &  1.05    & 6.9     & $1201^{+14}_{-8}$   & $123^{+14}_{-6}$   & $-0.08^{+0.07}_{-0.04}$ & $-0.03^{+0.06}_{-0.01}$ \\
NGC~7457* & B & 9.9    &  1.28    & 4.5     & $660 ^{+27}_{-39}$  & $61^{+15}_{-18}$   & $0.28^{+0.03}_{-0.08}$  & $0.31^{+0.08}_{-0.01}$ \\
\enddata
\tablecomments{Galaxies marked * are those with \m\ presented in the peer reviewed literature. Galaxies without data are those potentially contaminated by LINER  [Fe II]~$\lambda$8616 emission. The letters associated with the template stars used to make the fits are matched to the observing setups for the galaxies (Table~\ref{tab:sample}). A = 0\farcs1 slit, x1 binning; B = \farcs1, x2; C = 0\farcs2, x1; D = 0\farcs2, x2. The ranges given for the estimated LOSVD parameters are the 95\% bootstrap confidence bands.}
\end{deluxetable*}

\section{LOSVDs}

S/N alone cannot be used to quantify the quality of an estimated LOSVD. There have been many methods used to recover LOSVDs from galaxies using template spectra \citep[e.g.,][]{1974A&amp;A....31..129S,1977ApJ...212..326S,1989ApJ...344..613F,1990A&amp;A...229..441B,1992MNRAS.254..389R,1993MNRAS.264..712K,1993ApJ...407..525V,2005ApJS..160...76B}. The current trend in inferring the moments of LOSVDs has been to use a maximum penalized likelihood (MPL) method \citep[e.g.,][]{2003ApJ...596..903P,2004PASP..116..138C}. As it is known to perform well in the case of low S/N spectra, we use a well established MPL method here \citep{1997AJ....114..228M}.

In brief, the MPL approach uses a velocity grid to estimate the true LOSVD based on an optimal fit between a convolved stellar template and the galaxy spectrum. No implicit assumptions are made about the form of the LOSVD, and a penalty is applied to the fit based on the level of smoothness in the estimated LOSVD (where $\alpha$ is the smoothing factor). As $\alpha$ is increased, the estimated LOSVD tends toward a gaussian, therefore it is desirable to introduce as little smoothing as possible to limit the amount of bias in the estimated LOSVD. While the MPL approach estimates the LOSVD non-parametrically, the LOSVD is parameterized using the Gauss-Hermite series. The $h_3$ and $h_4$ coefficients of the Gauss-Hermite series (which are similar to skewness and kurtosis, respectively) can then be used to correct the estimated velocity dispersion, $\sigma_c = \sigma_0(1+\sqrt{6}h_4)$, and radial velocity, $v_c = v_0 + \sqrt{3}\sigma_0h_3$ \citep{1993ApJ...407..525V,1993MNRAS.265..213G}.

\begin{figure*}
\includegraphics*[width=480pt]{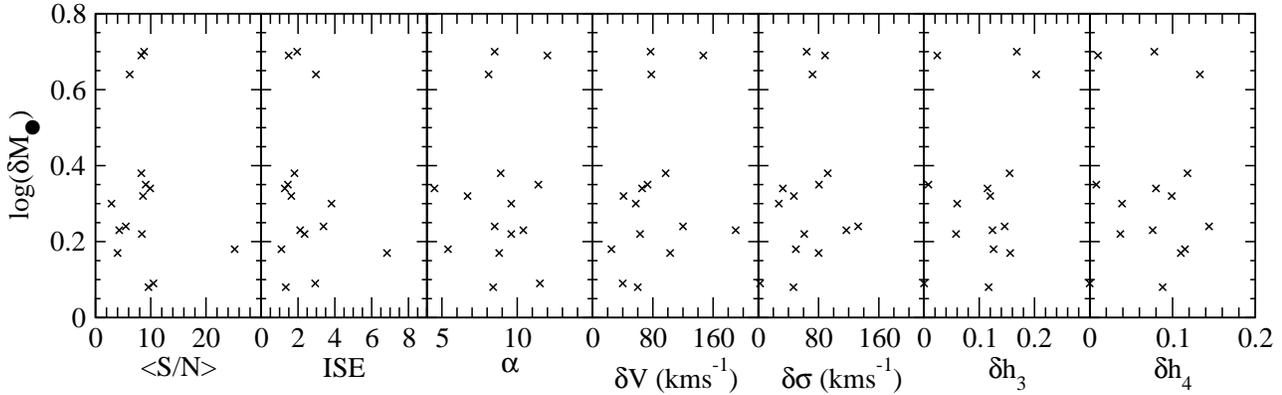}
\caption{Relationships between $<$S/N$>$ (the mean S/N in the central galaxy spectra), ISE (the integrated square error between the estimated LOSVD and the spectra), the MPL smoothing parameter ($\alpha$), the outputs from the MPL routine, and the uncertainty in \m\ ($\delta$\m).}
\label{fig:bhvs}
\end{figure*}

\begin{figure*}
\includegraphics*[width=480pt]{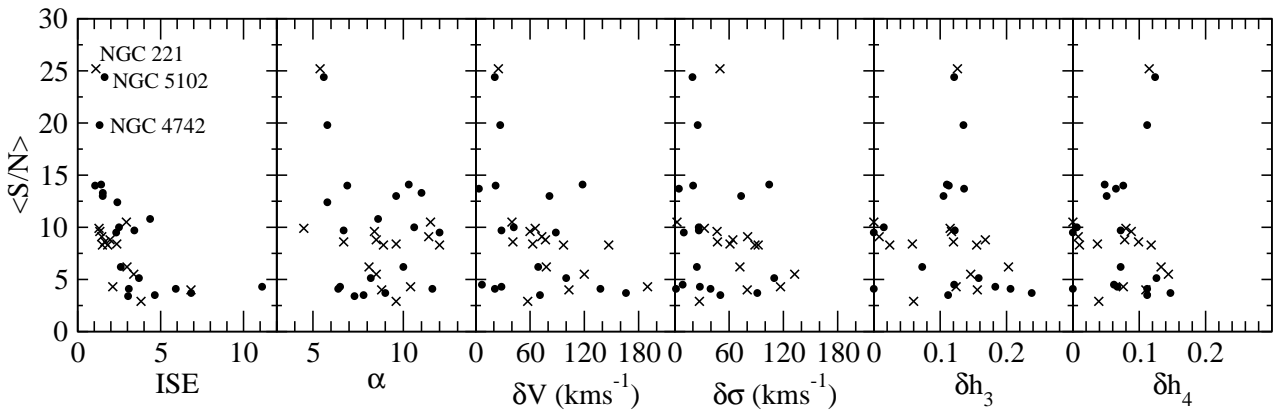}
\caption{Comparing the mean S/N in the spectra to the distributions of uncertainty in those data without previous \m\ estimates (filled circles) and those that do (crosses). Three clear S/N outliers are marked in the ISE panel.}
\label{fig:nobhvs}
\end{figure*}

Our MPL procedure for the STIS spectra was as follows. First, we match the observational setups for the template spectra and the galaxy spectra, i.e., the binning and slit width of the data to the four different templates. The optimum (lowest) value of the smoothing parameters was then determined by progressively increasing $\alpha$ in 0.1 increments until there was only a single stationary point in the estimated LOSVD at the radial velocity, i.e., $\alpha$ was slowly increased until the fluctuations from grid point to grid point disappeared; the estimated LOSVD was smooth, single, and not overly biased. The number of velocity grid points used was then adjusted to ensure the value was large enough to have no influence on the inferred moments of the LOSVD. The confidence in the LOSVD moments was then determined by a bootstrap. If the resulting confidence intervals did not fall either side of the optimal values, $\alpha$ was increased and the bootstrap repeated. 

\section{Results}\label{results}

Table~\ref{tab:results} presents the values of $\alpha$ used and the outputs from the MPL routine. It includes the integrated square error (ISE) between the estimated LOSVD and the observed spectra \citep{1997AJ....114..228M}. The Appendix presents the estimated LOSVDs with respect to the observed spectra, along with the spatial position of the STIS slit overlaid on the acquisition image. The Appendix also contains radial S/N profiles for each galaxy to demonstrate the spatial extent of the data.

In four cases (NGC~3608, NGC~4382, NGC~4552 and NGC~4564), the values of $h_3$ and $h_4$ have been driven to zero by large smoothing parameters ($\alpha\geq11.4)$. This is to be expected in such cases where the required smoothing parameter forces the LOSVD toward a gaussian. However, there is no relation between $\alpha$ and S/N. For example, NGC~1374, NGC~3607, NGC~3706 and NGC~4486A all have S/N$<$4 but $\alpha\leq9.6$). 

In six cases (those with only the S/N presented in Table~\ref{tab:results}) no estimates of the LOSVD were possible. Each of these galaxies are known to host low-ionization nuclear emission-line regions (LINERs), and we see a clear signature of [Fe II]~$\lambda$8616 in NGC~3031 (the highest S/N case) and NGC~3998. However, Fe II]~$\lambda$8616 emission is not detected in NGC~4278, NGC~4621 and NGC~7213. The STIS spectrum for NGC~4486 seems to be featureless, despite having a S/N much greater than members of the sample with well defined LOSVD. However, NGC~4486 is the dominant elliptical in the Virgo cluster with the most massive reported SBH in our sample. Compared to a less massive giant elliptical in the Virgo cluster, NGC~4649 where $\sigma_c=551^{35}_{-82}\kms$, an intrinsic value of $\sigma_c>550\kms$ is expected for NGC~4486. As demonstrated in the Appendix data for NGC~4649, when $\sigma_c$ becomes this large the signature of the absorption lines becomes very small - almost negligible. These absorption line signatures would be even less if $\sigma_c$ in NGC~4486 is intrinsically larger than 550\kms. However, it is the low number of constraining data points left in the template spectrum, once it has been broadened to $>$575\kms, that prevents an estimate of the LOSVD to be made. 

Of the remaining sample, 15 have peer reviewed \m\ estimates made using the spectra presented in the Appendix. The reference for NGC~4742 \citep{2002ApJ...574..740T} leads to an unpublished paper (Kramer et al. in prep), and the NGC~7332 reference \citep{2004ApJ...604L..89H} quotes a private communication. As a consequence, both of these galaxies are considered part of the `yet to be modeled' sample. As noted by \cite{2008PASA...25..167G}, the \m\ estimates quoted for NGC~4552 and NGC~4621 \citep{2008MNRAS.386.2242H} have been extracted from the figures of an IAU Symposium article \citep{2008IAUS..245..215C}. Therefore, both of these galaxies are also considered to be part of the 'yet to be modeled' sample. NGC~205 and NGC~598 only have upper limits to \m\ and are also excluded. These 15 galaxies allow us to examine whether there are any fundamental relationships between the uncertainty of the inferred moments of the LOSVDs and the \m\ estimates (Figure~\ref{fig:bhvs}). The uncertainty in the \m\ estimates presented in Table~\ref{tab:sample}, where $\delta$\m (the published range of acceptable values) is compared to the S/N, $\alpha$, the ISE, and the uncertainty in the inferred moments of the LOSVD, where $\delta\sigma=(\sigma_{max}-\sigma_{min})$ for example. Figure~\ref{fig:bhvs} demonstrates the uncertainty in \m\ is independent of the quality of the LOSVDs, and that the S/N$\leq10$ in all but one case (NGC~221). 

\begin{figure}
\plotone{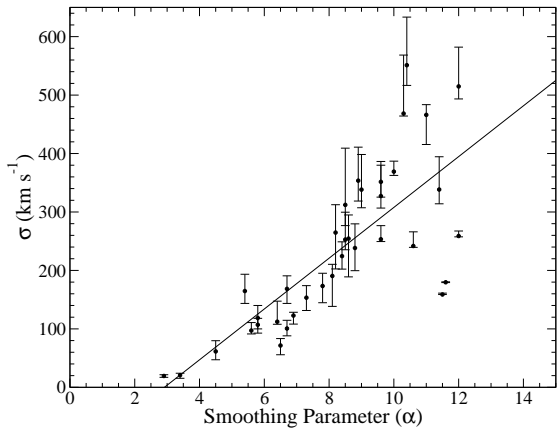}
\caption{The relationship between the estimates values of $\sigma$ and the MPL smoothing parameter. The solid black line marks the best fit relation. The correlation coefficient is 0.76, the slope is 43.5 (\kms/$\alpha$) and the intercept is -126.7~\kms.}
\label{fig:a-vs-sigma}
\end{figure}

To determine the suitability of the 19 remaining STIS spectra for estimates of SBH masses, we can compare the distributions of uncertainty in the inferred moments of the LOSVDs for these galaxies with the 15 galaxies from before (Figure~\ref{fig:nobhvs}). Immediately apparent are two galaxies with high S/N and low uncertainties (NGC~5102 and NGC~4742). As noted for Figure~\ref{fig:bhvs}, NGC~221 also lies in the same high S/N-low uncertainty region of Figure~\ref{fig:nobhvs}. However, aside from these three cases, there are no statistically significant differences between those data already used to estimate \m\ and those that have not. 

\section{Discussions}\label{disc}

From these data, NGC~5102 and NGC~4742 have the highest potential for new high quality estimates of \m\ using stellar dynamics. Both show high S/N and low uncertainties in their LOSVDS. In addition, they have spatially extended data with good S/N (that may provide excellent constraints on the bulge gravitational potential), and very prominent CaII absorption. This is not the case for the majority of the remaining spectra however. 

To the unaided eye, many galaxies in the sample do not exhibit obvious absorption-line features, including some galaxies for which estimates of \m\ have already been published. The S/N levels are also all very low. This demonstrates the advantages of MPL; even with data such as these relative uncertainties of $\sim10\%$ in $v_c$ and $\sim20\%$ in $\sigma_c$ can be recovered. The quality of these spectra will, however, deteriorate as spatial positions either side of the S/N peaks are sampled to constrain the gravitational potential. Regardless, \m\ estimates with relative uncertainties of $\sim70\pm30\%$ mostly from \cite{2003ApJ...583...92G} and \cite{2009ApJ...695.1577G}, have been extracted from these data, and there is nothing to indicate that the remaining 19 galaxies would not yield many useful estimates of \m. It is not unexpected that these galaxies have yet to have \m\ estimates attempted; in most cases these data have not be collected with the express purposes of determining \m. 

We find no correlation between S/N and the inferred moments of the LOSVD. Therefore, there is no simple way to determine the relative uncertainty of any LOSVD, and as a consequence \m, based solely on an estimated S/N from an exposure time calculator. Principle component analysis may reveal a deeper relationship between S/N and the relative uncertainties in the estimated LOSVD, but our aim here is to simply determine the best candidates in the archive for further \m\ modeling. However, there are only a very limited number of suitable stellar templates with which to extract the LOSVDs. Consequently, it is possible that some of the scatter seen in the uncertainties may be due to various levels of template mismatching between the galaxy spectra and standard stars. Increasing the variety of standard stars observed with STIS using a multitude of common instrument configurations would be beneficial in this case. In addition, some improvements may result from using the extensive stellar libraries of \cite{2001MNRAS.326..959C} as kinematic templates, and by using linear combinations of the stellar template spectra currently available that may better match the galaxy spectra. 

Non-gravitational kinematics, nuclear star clusters, multiple nuclei, dust, spatially offset SBHs, triaxial potentials, dark matter halos and mathematical degeneracy of solutions \citep{2004ApJ...602...66V} may all play non-trivial roles in the complexity of estimating $\delta$\m\ , so the lack of correlation between $\delta$\m\ and the estimated LOSVD parameters in Figure~\ref{fig:bhvs} is not surprising. Indeed, many of these factors determine the complexity of the observed absorption features and consequently the precision with which the LOSVD can be estimated. As a result, the lack of correlation seen between S/N and the estimated LOSVD parameters in Figure~\ref{fig:nobhvs} is also not surprising. For example, NGC 205 has a peak S/N of 4.5, a small smoothing parameter (3.4), but easily detected CaII absorption. The absolute uncertainty in the radial velocity is 6.8 \kms (3\% relative). Conversely, NGC 221 has a peak S/N of 25.22, $\alpha$=5.4, with less easily detected CaII absorption. Here the absolute uncertainty in radial velocity is 25 \kms (8\% relative). Both these objects are well known dwarf galaxies, but the differences in the relative uncertainty in the LOSVDs are not unexpected when it is considered that NGC 205 contains a well defined compact nuclear star cluster. However, as expected and as demonstrated by Figure~\ref{fig:a-vs-sigma}, there is a strong correlation between absolute value of $\sigma$ and the smoothing parameter. 

\begin{deluxetable*}{lccc|lccc}
\tabletypesize{\footnotesize}
\tablecaption{Galaxies with Multiple Peer Reviewed \m\ Estimates\label{tab:twobhs}}
\tablewidth{0pt}
\tablehead{
\colhead{Name}&
\colhead{$\log M_{\bullet{1}}$}&
\colhead{$\log M_{\bullet{2}}$}&
\colhead{$\Delta\log M_\bullet$}&
\colhead{Name}&
\colhead{$\log M_{\bullet{1}}$}&
\colhead{$\log M_{\bullet{2}}$}&
\colhead{$\Delta\log M_\bullet$}\\
}
\startdata
IC~1459		& $^a8.54_{-0.30}^{+0.30}$ (g, 1) 		& $9.41_{-0.23}^{+0.16}$ (s, 1)		& i, 0.87	& NGC~3998$^b$	& $8.34_{-0.64}^{+0.28}$ (g, 19) 		& $8.91_{-0.12}^{+0.09}$ (s, 20)	& i, 0.57	\\
IC~2560		& $^a6.45_{-0.30}^{+0.30}$ (m, 2)		& $6.54_{-0.06}^{+0.06}$ (m, 3)	& c, 0.09	& NGC~4151		& $7.48_{-0.57}^{+0.10}$ (g, 13)		& $7.60-7.70$ (s, 21) 			& i, 0.17	\\
NGC~224$^b$	& $7.54-7.93$ (s, 4) 					& $8.15_{-0.11}^{+0.21}$ (s, 5)		& i, 0.42	& cNGC~4258		& $7.591_{-0.001}^{+0.001}$ (m, 22)	& $7.52_{-0.04}^{+0.03}$ (g, 23)	& i, 0.07	\\
NGC~598$^b$	& $<3.18$ (s, 6) 					& $<3.48$ (s, 7)				& n/a		& NGC~4350		& $8.18-8.99$ (s\&g, 24) 				& $^a8.90_{-0.30}^{+0.30}$ (s, 25)	& c, 0.32	\\
NGC~1023	& $7.78_{-0.12}^{+0.09}$ (s, 8) 		& $7.59_{-0.05}^{+0.04}$ (s, 8)		& i, 0.19 	& NGC~4374		& $9.18_{-0.23}^{0.23}$ (g, 26)			& $^a8.60_{-0.30}^{0.30}$ (g, 27)	& i, 0.58	\\
NGC~2960	& $7.04_{-0.14}^{+0.11}$ (m, 9) 		& $7.06_{-0.02}^{+0.02}$ (m, 10)	& c, 0.02	& NGC~4486$^b$	& $9.53_{-0.15}^{+0.11}$ (g, 28) 		& $9.82_{-0.03}^{+0.03}$ (s, 29)	& i, 0.29	\\
NGC~3079	& $^a6.00_{-0.30}^{+0.30}$ (m, 11) 		& $^a6.30_{-0.30}^{+0.30}$ (m, 12)	& c, 0.30	& NGC~4649$^b$	& $9.30_{-0.15}^{+0.08}$ (s, 15) 		& $9.65_{-0.11}^{+0.09}$ (s, 30)	& i, 0.35	\\
NGC~3227	& $7.30_{-0.10}^{+0.18}$ (g, 13)		& $6.85-7.30$ (s, 14) 			& c, 0.27	& NGC~5128		& $7.65_{-0.11}^{+0.14}$ (g, 31) 		& $7.74_{-0.34}^{+0.19}$ (s, 32)	& c, 0.09	\\
NGC~3377	& $8.00_{-0.05}^{+0.28}$ (s, 15) 		& $7.85_{-0.55}^{+0.19}$ (s, 16)	& c, 0.15	& UGC~3789		& $^a6.95_{-0.30}^{+0.30}$ (m, 33) 		& $7.02_{-0.02}^{+0.02}$ (m, 34)	& c, 0.07	\\
NGC~3379	& $8.15_{-0.55}^{+0.55}$ (s, 17) 		& $8.60_{-0.12}^{+0.10}$ (s, 18)	& c, 0.45	
\enddata
\tablecomments{Estimates of \m\ from more than one peer reviewed source. The estimated \m\ method codes are: g = gas kinematics, m = megamasers, s = stellar dynamics. $\Delta\log M_\bullet$ is the difference between the two quoted \m\ estimates, and the codes reference whether the uncertainties in the two estimates are c = consistent or i = inconsistent. Our consistency criteria is that either $dM_{\bullet{1}} / \Delta\log M_\bullet > 1.0$ or $dM_{\bullet{2}} / \Delta\log M_\bullet > 1.0$. Note that there has yet to be a standard \m\ uncertainty adopted by the community, and that some authors choose 1$\sigma$, some choose 3$\sigma$, and some do not state how their uncertainties were estimated.\\
$^a$ indicates a factor of 2 uncertainty has been assumed for \m\ because no estimate of the uncertainty could be found in the literature. \\
$^b$ indicates galaxies also in the Table~\ref{tab:sample} sample. \\
{\bf References:}
(1) \cite{2002ApJ...578..787C}; 
(2) \cite{2001PASJ...53..215I}; 
(3) \cite{2012arXiv1207.6820Y}; 
(4) \cite{2001A&A...371..409B}; 
(5) \cite{2005ApJ...631..280B}; 
(6) \cite{2001AJ....122.2469G};  
(7) \cite{2001Sci...293.1116M}; 
(8) \cite{2001ApJ...550...75B}; 
(9) \cite{2002A&amp;A...394L..23H}; 
(10) \cite{2011ApJ...727...20K}; 
(11) \cite{2004PASJ...56..605Y}; 
(12) \cite{2005ApJ...618..618K}; 
(13) \cite{2008ApJS..174...31H}; 
(14) \cite{2006ApJ...646..754D}; 
(15) \cite{2003ApJ...583...92G}; 
(16) \cite{2004A&amp;A...415..889C}; 
(17) \cite{2006MNRAS.370..559S}; 
(18) \cite{2010MNRAS.401.1770V}; 
(19) \cite{2006A&A...460..439D}; 
(20) \cite{2012ApJ...753...79W}; 
(21) \cite{2007ApJ...670..105O}; 
(22) \cite{1995Natur.373..127M}; 
(23) \cite{2009ApJ...693..946S}; 
(24) \cite{2001MNRAS.320..124P}; 
(25) \cite{1998A&amp;A...334..805L}; 
(26) \cite{1998ApJ...492L.111B}; 
(27) \cite{2001MNRAS.323..831M}; 
(28) \cite{1997ApJ...489..579M}; 
(29) \cite{2011ApJ...729..119G}; 
(30) \cite{2010ApJ...711..484S}; 
(31) \cite{2007ApJ...671.1329N}; 
(32) \cite{2009MNRAS.394..660C}; 
(33) \cite{2008ApJ...678...96B}; 
(34) \cite{2011ApJ...727...20K}.
}
\end{deluxetable*}

As the number of different \m\ estimates using different approaches increases, the consistency between \m\ techniques can be examined in those galaxies where more than one method has been applied. Table~\ref{tab:twobhs} presents 19 galaxies with 2 \m\ estimates each; 9 from gas kinematics (GK), 9 from megamasers (MM), 20 from stellar dynamics (SD). NGC~221, NGC~4486 and NGC~5128 have more than 2 \m\ estimates; see \cite{2005ApJ...631..280B}, \cite{2011ApJ...729..119G} and \cite{2010PASA...27..449N} for more details. NGC~598 only has published upper limits and is excluded from further discussion. Nine galaxies have consistent \m\ estimates; 4 are MM to MM, 3 are SD to SD, and 2 are GK to SD comparisons. Nine galaxies have inconsistent \m\ estimates; 1 MM to GK, 3 SD to SD, 1 GK to GK, and 4 GK to SD. Excluding the MMs, only 38\% of \m\ estimates are consistent; there is a dispersion of 0.3 dex about the one-to-one relation. However, as demonstrated by the differences between galaxies with two separate SD and GK estimates, there still appears to be ample motivation for refining current modeling techniques (e.g., using the CO band-head and adaptive optics to get {\it HST}-like spatial resolution, \citealt{2011ApJ...729..119G}), or perhaps pursuing some new ones like spectroastrometry of nuclear gas disks \citep{2012arXiv1211.0943G} and the kinematics of molecular gas \citep{2013Natur.494..328D}.

For the six galaxies in this sample that have AGN contamination it may be possible to mask out, or model, the [FeII] $8616\lambda$ emission. This was achieved by \cite{2012ApJ...753...79W}, for example, with NGC~3998. While the data that dominated the \m\ estimate in this case was from Keck adaptive optics measurements of the CO band-heads, the authors were able to successfully recover the LOSVDs of this galaxy from the STIS spectra presented here. 

\section{Conclusions}\label{cons}

We have identified galaxies in the STIS archive with spectra of sufficient quality that they could be used for stellar-dynamical estimates of \m. These data were retrieved, 
co-aligned, combined, and their radial S/N calculated. The LOSVDs were also fitted and the resulting data are presented in this atlas. In addition to the 15 galaxies that have already been modeled in this way, there are another 19 for which the data quality is comparable. Two galaxies are particularly noteworthy: NGC 5102 and NGC 4742 both have spectra of relatively high S/N and correspondingly well-determined LOSVDs. NGC 5055 is also an interesting prospect, since it has already been modeled using gas kinematics, and a stellar dynamical model might help explain why gas- and stellar dynamical mass estimates are so often discrepant.

The uncertainties in the inferred moments of the LOSVDs derived from the central STIS spectra show no correlations with S/N, i.e., there appears to be no way to determine the accuracy with which an estimate of \m\ can be made simply from S/N. This highlights the difficulty in estimating precise values of \m\ because, in addition to the need for intricate data sets, the complicated methods being used give \m\ uncertainties that are independent of the measurement errors. However, there may be excessive scatter in the estimated LOSVDs due to template mismatching from the limited number of stellar templates observed using the appropriate instrument configurations. Further STIS observations of a range of stellar templates using slit widths of 0\farcs1 and 0\farcs2, un-binned and x2 binned data, and the appropriate contemporaneous fringe flats, may provide a significant improvement to the accuracy of the estimated LOSVDs. 

\acknowledgments

We would like to thank the anonymous referee for their excellent comments that improved the quality of this manuscript. 
DB wishes to thank K. Azalee Bostroem, Brian York, Jerry Kriss and Phil Hodge for their help and patience with the multiple questions regarding the reduction of the STIS spectra. Support for Proposal number HST-AR-10935.01 was provided by NASA through a grant from the Space Telescope Science Institute, which is operated by the Association of Universities for Research in Astronomy, Incorporated, under NASA contract NAS5-26555. This research has made use of the NASA/IPAC Extragalactic Database (NED) which is operated by the Jet Propulsion Laboratory, California Institute of Technology, under contract with the National Aeronautics and Space Administration. We acknowledge the usage of the HyperLeda database (http://leda.univ-lyon1.fr).

\bibliographystyle{apj}
\bibliography{batcheldor}

\begin{thebibliography}{105}
\expandafter\ifx\csname natexlab\endcsname\relax\def\natexlab#1{#1}\fi

\bibitem[{{Adams} {et~al.}(2003){Adams}, {Graff}, {Mbonye}, \&
  {Richstone}}]{2003ApJ...591..125A}
{Adams}, F.~C., {Graff}, D.~S., {Mbonye}, M., \& {Richstone}, D.~O. 2003, \apj,
  591, 125

\bibitem[{{Bacon} {et~al.}(2001){Bacon}, {Emsellem}, {Combes}, {Copin},
  {Monnet}, \& {Martin}}]{2001A&A...371..409B}
{Bacon}, R., {Emsellem}, E., {Combes}, F., {Copin}, Y., {Monnet}, G., \&
  {Martin}, P. 2001, \aap, 371, 409

\bibitem[{{Baes} {et~al.}(2003){Baes}, {Buyle}, {Hau}, \&
  {Dejonghe}}]{2003MNRAS.341L..44B}
{Baes}, M., {Buyle}, P., {Hau}, G.~K.~T., \& {Dejonghe}, H. 2003, \mnras, 341,
  L44

\bibitem[{{Batcheldor}(2010)}]{2010ApJ...711L.108B}
{Batcheldor}, D. 2010, \apjl, 711, L108

\bibitem[{{Batcheldor} {et~al.}(2005){Batcheldor}, {Axon}, {Merritt}, {Hughes},
  {Marconi}, {Binney}, {Capetti}, {Merrifield}, {Scarlata}, \&
  {Sparks}}]{2005ApJS..160...76B}
{Batcheldor}, D., {Axon}, D., {Merritt}, D., {Hughes}, M.~A., {Marconi}, A.,
  {Binney}, J., {Capetti}, A., {Merrifield}, M., {Scarlata}, C., \& {Sparks},
  W. 2005, \apjs, 160, 76

\bibitem[{{Bender}(1990)}]{1990A&amp;A...229..441B}
{Bender}, R. 1990, \aap, 229, 441

\bibitem[{{Bender} {et~al.}(2005){Bender}, {Kormendy}, {Bower}, {Green},
  {Thomas}, {Danks}, {Gull}, {Hutchings}, {Joseph}, {Kaiser}, {Lauer},
  {Nelson}, {Richstone}, {Weistrop}, \& {Woodgate}}]{2005ApJ...631..280B}
{Bender}, R., {Kormendy}, J., {Bower}, G., {Green}, R., {Thomas}, J., {Danks},
  A.~C., {Gull}, T., {Hutchings}, J.~B., {Joseph}, C.~L., {Kaiser}, M.~E.,
  {Lauer}, T.~R., {Nelson}, C.~H., {Richstone}, D., {Weistrop}, D., \&
  {Woodgate}, B. 2005, \apj, 631, 280

\bibitem[{{Blais-Ouellette} {et~al.}(2004){Blais-Ouellette}, {Amram},
  {Carignan}, \& {Swaters}}]{2004A&A...420..147B}
{Blais-Ouellette}, S., {Amram}, P., {Carignan}, C., \& {Swaters}, R. 2004,
  \aap, 420, 147

\bibitem[{{Bower} {et~al.}(2001){Bower}, {Green}, {Bender}, {Gebhardt},
  {Lauer}, {Magorrian}, {Richstone}, {Danks}, {Gull}, {Hutchings}, {Joseph},
  {Kaiser}, {Weistrop}, {Woodgate}, {Nelson}, \&
  {Malumuth}}]{2001ApJ...550...75B}
{Bower}, G.~A., {Green}, R.~F., {Bender}, R., {Gebhardt}, K., {Lauer}, T.~R.,
  {Magorrian}, J., {Richstone}, D.~O., {Danks}, A., {Gull}, T., {Hutchings},
  J., {Joseph}, C., {Kaiser}, M.~E., {Weistrop}, D., {Woodgate}, B., {Nelson},
  C., \& {Malumuth}, E.~M. 2001, \apj, 550, 75

\bibitem[{{Bower} {et~al.}(1998){Bower}, {Green}, {Danks}, {Gull}, {Heap},
  {Hutchings}, {Joseph}, {Kaiser}, {Kimble}, {Kraemer}, {Weistrop}, {Woodgate},
  {Lindler}, {Hill}, {Malumuth}, {Baum}, {Sarajedini}, {Heckman}, {Wilson}, \&
  {Richstone}}]{1998ApJ...492L.111B}
{Bower}, G.~A., {Green}, R.~F., {Danks}, A., {Gull}, T., {Heap}, S.,
  {Hutchings}, J., {Joseph}, C., {Kaiser}, M.~E., {Kimble}, R., {Kraemer}, S.,
  {Weistrop}, D., {Woodgate}, B., {Lindler}, D., {Hill}, R.~S., {Malumuth},
  E.~M., {Baum}, S., {Sarajedini}, V., {Heckman}, T.~M., {Wilson}, A.~S., \&
  {Richstone}, D.~O. 1998, \apjl, 492, L111

\bibitem[{{Braatz} \& {Gugliucci}(2008)}]{2008ApJ...678...96B}
{Braatz}, J.~A. \& {Gugliucci}, N.~E. 2008, \apj, 678, 96

\bibitem[{{Cappellari} {et~al.}(2008){Cappellari}, {Bacon}, {Davies}, {de
  Zeeuw}, {Emsellem}, {Falc{\'o}n-Barroso}, {Krajnovi{\'c}}, {Kuntschner},
  {McDermid}, {Peletier}, {Sarzi}, {van den Bosch}, \& {van de
  Ven}}]{2008IAUS..245..215C}
{Cappellari}, M., {Bacon}, R., {Davies}, R.~L., {de Zeeuw}, P.~T., {Emsellem},
  E., {Falc{\'o}n-Barroso}, J., {Krajnovi{\'c}}, D., {Kuntschner}, H.,
  {McDermid}, R.~M., {Peletier}, R.~F., {Sarzi}, M., {van den Bosch}, R.~C.~E.,
  \& {van de Ven}, G. 2008, in IAU Symposium, Vol. 245, IAU Symposium, ed.
  M.~{Bureau}, E.~{Athanassoula}, \& B.~{Barbuy}, 215--218

\bibitem[{{Cappellari} \& {Emsellem}(2004)}]{2004PASP..116..138C}
{Cappellari}, M. \& {Emsellem}, E. 2004, \pasp, 116, 138

\bibitem[{{Cappellari} {et~al.}(2009){Cappellari}, {Neumayer}, {Reunanen}, {van
  der Werf}, {de Zeeuw}, \& {Rix}}]{2009MNRAS.394..660C}
{Cappellari}, M., {Neumayer}, N., {Reunanen}, J., {van der Werf}, P.~P., {de
  Zeeuw}, P.~T., \& {Rix}, H.-W. 2009, \mnras, 394, 660

\bibitem[{{Cappellari} {et~al.}(2002){Cappellari}, {Verolme}, {van der Marel},
  {Kleijn}, {Illingworth}, {Franx}, {Carollo}, \& {de
  Zeeuw}}]{2002ApJ...578..787C}
{Cappellari}, M., {Verolme}, E.~K., {van der Marel}, R.~P., {Kleijn}, G.~A.~V.,
  {Illingworth}, G.~D., {Franx}, M., {Carollo}, C.~M., \& {de Zeeuw}, P.~T.
  2002, \apj, 578, 787

\bibitem[{{Cattaneo} {et~al.}(2005){Cattaneo}, {Blaizot}, {Devriendt}, \&
  {Guiderdoni}}]{2005MNRAS.364..407C}
{Cattaneo}, A., {Blaizot}, J., {Devriendt}, J., \& {Guiderdoni}, B. 2005,
  \mnras, 364, 407

\bibitem[{{Cenarro} {et~al.}(2001){Cenarro}, {Cardiel}, {Gorgas}, {Peletier},
  {Vazdekis}, \& {Prada}}]{2001MNRAS.326..959C}
{Cenarro}, A.~J., {Cardiel}, N., {Gorgas}, J., {Peletier}, R.~F., {Vazdekis},
  A., \& {Prada}, F. 2001, \mnras, 326, 959

\bibitem[{{Ciotti}(2008)}]{2008arXiv0808.1349C}
{Ciotti}, L. 2008, ArXiv e-prints

\bibitem[{{Ciotti} \& {van Albada}(2001)}]{2001ApJ...552L..13C}
{Ciotti}, L. \& {van Albada}, T.~S. 2001, \apjl, 552, L13

\bibitem[{{Copin} {et~al.}(2004){Copin}, {Cretton}, \&
  {Emsellem}}]{2004A&amp;A...415..889C}
{Copin}, Y., {Cretton}, N., \& {Emsellem}, E. 2004, \aap, 415, 889

\bibitem[{{Cretton} {et~al.}(1999){Cretton}, {de Zeeuw}, {van der Marel}, \&
  {Rix}}]{1999ApJS..124..383C}
{Cretton}, N., {de Zeeuw}, P.~T., {van der Marel}, R.~P., \& {Rix}, H.-W. 1999,
  \apjs, 124, 383

\bibitem[{{Davies} {et~al.}(2006){Davies}, {Thomas}, {Genzel}, {S{\'a}nchez},
  {Tacconi}, {Sternberg}, {Eisenhauer}, {Abuter}, {Saglia}, \&
  {Bender}}]{2006ApJ...646..754D}
{Davies}, R.~I., {Thomas}, J., {Genzel}, R., {S{\'a}nchez}, F.~M., {Tacconi},
  L.~J., {Sternberg}, A., {Eisenhauer}, F., {Abuter}, R., {Saglia}, R., \&
  {Bender}, R. 2006, \apj, 646, 754

\bibitem[{{Davis} {et~al.}(2013){Davis}, {Bureau}, {Cappellari}, {Sarzi}, \&
  {Blitz}}]{2013Natur.494..328D}
{Davis}, T.~A., {Bureau}, M., {Cappellari}, M., {Sarzi}, M., \& {Blitz}, L.
  2013, \nat, 494, 328

\bibitem[{{de Francesco} {et~al.}(2006){de Francesco}, {Capetti}, \&
  {Marconi}}]{2006A&A...460..439D}
{de Francesco}, G., {Capetti}, A., \& {Marconi}, A. 2006, \aap, 460, 439

\bibitem[{{Devereux} {et~al.}(2003){Devereux}, {Ford}, {Tsvetanov}, \&
  {Jacoby}}]{2003AJ....125.1226D}
{Devereux}, N., {Ford}, H., {Tsvetanov}, Z., \& {Jacoby}, G. 2003, \aj, 125,
  1226

\bibitem[{{Emsellem} {et~al.}(1999){Emsellem}, {Dejonghe}, \&
  {Bacon}}]{1999MNRAS.303..495E}
{Emsellem}, E., {Dejonghe}, H., \& {Bacon}, R. 1999, \mnras, 303, 495

\bibitem[{{Ferrarese}(2002)}]{2002ApJ...578...90F}
{Ferrarese}, L. 2002, \apj, 578, 90

\bibitem[{{Ferrarese} {et~al.}(1996){Ferrarese}, {Ford}, \&
  {Jaffe}}]{1996ApJ...470..444F}
{Ferrarese}, L., {Ford}, H.~C., \& {Jaffe}, W. 1996, \apj, 470, 444

\bibitem[{{Ferrarese} \& {Merritt}(2000)}]{2000ApJ...539L...9F}
{Ferrarese}, L. \& {Merritt}, D. 2000, \apjl, 539, L9

\bibitem[{{Franx} {et~al.}(1989){Franx}, {Illingworth}, \&
  {Heckman}}]{1989ApJ...344..613F}
{Franx}, M., {Illingworth}, G., \& {Heckman}, T. 1989, \apj, 344, 613

\bibitem[{{Gebhardt} {et~al.}(2011){Gebhardt}, {Adams}, {Richstone}, {Lauer},
  {Faber}, {G{\"u}ltekin}, {Murphy}, \& {Tremaine}}]{2011ApJ...729..119G}
{Gebhardt}, K., {Adams}, J., {Richstone}, D., {Lauer}, T.~R., {Faber}, S.~M.,
  {G{\"u}ltekin}, K., {Murphy}, J., \& {Tremaine}, S. 2011, \apj, 729, 119

\bibitem[{{Gebhardt} {et~al.}(2000){Gebhardt}, {Bender}, {Bower}, {Dressler},
  {Faber}, {Filippenko}, {Green}, {Grillmair}, {Ho}, {Kormendy}, {Lauer},
  {Magorrian}, {Pinkney}, {Richstone}, \& {Tremaine}}]{2000ApJ...539L..13G}
{Gebhardt}, K., {Bender}, R., {Bower}, G., {Dressler}, A., {Faber}, S.~M.,
  {Filippenko}, A.~V., {Green}, R., {Grillmair}, C., {Ho}, L.~C., {Kormendy},
  J., {Lauer}, T.~R., {Magorrian}, J., {Pinkney}, J., {Richstone}, D., \&
  {Tremaine}, S. 2000, \apjl, 539, L13

\bibitem[{{Gebhardt} {et~al.}(2001){Gebhardt}, {Lauer}, {Kormendy}, {Pinkney},
  {Bower}, {Green}, {Gull}, {Hutchings}, {Kaiser}, {Nelson}, {Richstone}, \&
  {Weistrop}}]{2001AJ....122.2469G}
{Gebhardt}, K., {Lauer}, T.~R., {Kormendy}, J., {Pinkney}, J., {Bower}, G.~A.,
  {Green}, R., {Gull}, T., {Hutchings}, J.~B., {Kaiser}, M.~E., {Nelson},
  C.~H., {Richstone}, D., \& {Weistrop}, D. 2001, \aj, 122, 2469

\bibitem[{{Gebhardt} {et~al.}(2003){Gebhardt}, {Richstone}, {Tremaine},
  {Lauer}, {Bender}, {Bower}, {Dressler}, {Faber}, {Filippenko}, {Green},
  {Grillmair}, {Ho}, {Kormendy}, {Magorrian}, \&
  {Pinkney}}]{2003ApJ...583...92G}
{Gebhardt}, K., {Richstone}, D., {Tremaine}, S., {Lauer}, T.~R., {Bender}, R.,
  {Bower}, G., {Dressler}, A., {Faber}, S.~M., {Filippenko}, A.~V., {Green},
  R., {Grillmair}, C., {Ho}, L.~C., {Kormendy}, J., {Magorrian}, J., \&
  {Pinkney}, J. 2003, \apj, 583, 92

\bibitem[{{Gebhardt} \& {Thomas}(2009)}]{2009ApJ...700.1690G}
{Gebhardt}, K. \& {Thomas}, J. 2009, \apj, 700, 1690

\bibitem[{{Gerhard}(1993)}]{1993MNRAS.265..213G}
{Gerhard}, O.~E. 1993, \mnras, 265, 213

\bibitem[{{Gnerucci} {et~al.}(2012){Gnerucci}, {Marconi}, {Capetti}, {Axon}, \&
  {Robinson}}]{2012arXiv1211.0943G}
{Gnerucci}, A., {Marconi}, A., {Capetti}, A., {Axon}, D.~J., \& {Robinson}, A.
  2012, ArXiv e-prints

\bibitem[{{Goudfrooij} {et~al.}(1998){Goudfrooij}, {Bohlin}, {Walsh}, \&
  {Baum}}]{1998stis.rept...19G}
{Goudfrooij}, P., {Bohlin}, R.~C., {Walsh}, J.~R., \& {Baum}, S.~A. 1998, {STIS
  Near-IR Fringing. II. Basics and Use of Contemporaneous Flats for
  Spectroscopy of Point Sources (Rev. A)}, Tech. rep., Space Telescope Science
  Institute

\bibitem[{{Goudfrooij} \& {Christensen}(1998)}]{1998stis.rept...29G}
{Goudfrooij}, P. \& {Christensen}, J.~A. 1998, {STIS Near-IR Fringing. III. A
  Tutorial on the Use of the IRAF Tasks}, Tech. rep., Space Telescope Science
  Institute

\bibitem[{{Graham}(2008)}]{2008PASA...25..167G}
{Graham}, A.~W. 2008, Publications of the Astronomical Society of Australia,
  25, 167

\bibitem[{{Graham} {et~al.}(2001){Graham}, {Erwin}, {Caon}, \&
  {Trujillo}}]{2001ApJ...563L..11G}
{Graham}, A.~W., {Erwin}, P., {Caon}, N., \& {Trujillo}, I. 2001, \apjl, 563,
  L11

\bibitem[{{Graham} \& {Li}(2009)}]{2009ApJ...698..812G}
{Graham}, A.~W. \& {Li}, I. 2009, \apj, 698, 812

\bibitem[{{Graham} {et~al.}(2011){Graham}, {Onken}, {Athanassoula}, \&
  {Combes}}]{2011MNRAS.412.2211G}
{Graham}, A.~W., {Onken}, C.~A., {Athanassoula}, E., \& {Combes}, F. 2011,
  \mnras, 412, 2211

\bibitem[{{Greene} {et~al.}(2010){Greene}, {Peng}, {Kim}, {Kuo}, {Braatz},
  {Violette Impellizzeri}, {Condon}, {Lo}, {Henkel}, \&
  {Reid}}]{2010ApJ...721...26G}
{Greene}, J.~E., {Peng}, C.~Y., {Kim}, M., {Kuo}, C.-Y., {Braatz}, J.~A.,
  {Violette Impellizzeri}, C.~M., {Condon}, J.~J., {Lo}, K.~Y., {Henkel}, C.,
  \& {Reid}, M.~J. 2010, \apj, 721, 26

\bibitem[{{G{\"u}ltekin} {et~al.}(2009){G{\"u}ltekin}, {Richstone}, {Gebhardt},
  {Lauer}, {Pinkney}, {Aller}, {Bender}, {Dressler}, {Faber}, {Filippenko},
  {Green}, {Ho}, {Kormendy}, \& {Siopis}}]{2009ApJ...695.1577G}
{G{\"u}ltekin}, K., {Richstone}, D.~O., {Gebhardt}, K., {Lauer}, T.~R.,
  {Pinkney}, J., {Aller}, M.~C., {Bender}, R., {Dressler}, A., {Faber}, S.~M.,
  {Filippenko}, A.~V., {Green}, R., {Ho}, L.~C., {Kormendy}, J., \& {Siopis},
  C. 2009, \apj, 695, 1577

\bibitem[{{H{\"a}ring} \& {Rix}(2004)}]{2004ApJ...604L..89H}
{H{\"a}ring}, N. \& {Rix}, H. 2004, \apjl, 604, L89

\bibitem[{{Heckman} {et~al.}(2004){Heckman}, {Kauffmann}, {Brinchmann},
  {Charlot}, {Tremonti}, \& {White}}]{2004ApJ...613..109H}
{Heckman}, T.~M., {Kauffmann}, G., {Brinchmann}, J., {Charlot}, S., {Tremonti},
  C., \& {White}, S.~D.~M. 2004, \apj, 613, 109

\bibitem[{{Henkel} {et~al.}(2002){Henkel}, {Braatz}, {Greenhill}, \&
  {Wilson}}]{2002A&amp;A...394L..23H}
{Henkel}, C., {Braatz}, J.~A., {Greenhill}, L.~J., \& {Wilson}, A.~S. 2002,
  \aap, 394, L23

\bibitem[{{Hicks} \& {Malkan}(2008)}]{2008ApJS..174...31H}
{Hicks}, E.~K.~S. \& {Malkan}, M.~A. 2008, \apjs, 174, 31

\bibitem[{{Hu}(2008)}]{2008MNRAS.386.2242H}
{Hu}, J. 2008, \mnras, 386, 2242

\bibitem[{{Hughes} {et~al.}(2003){Hughes}, {Alonso-Herrero}, {Axon},
  {Scarlata}, {Atkinson}, {Batcheldor}, {Binney}, {Capetti}, {Carollo},
  {Dressel}, {Gerssen}, {Macchetto}, {Maciejewski}, {Marconi}, {Merrifield},
  {Ruiz}, {Sparks}, {Stiavelli}, {Tsvetanov}, \& {van der
  Marel}}]{2003AJ....126..742H}
{Hughes}, M.~A., {Alonso-Herrero}, A., {Axon}, D., {Scarlata}, C., {Atkinson},
  J., {Batcheldor}, D., {Binney}, J., {Capetti}, A., {Carollo}, C.~M.,
  {Dressel}, L., {Gerssen}, J., {Macchetto}, D., {Maciejewski}, W., {Marconi},
  A., {Merrifield}, M., {Ruiz}, M., {Sparks}, W., {Stiavelli}, M., {Tsvetanov},
  Z., \& {van der Marel}, R. 2003, \aj, 126, 742

\bibitem[{{Ishihara} {et~al.}(2001){Ishihara}, {Nakai}, {Iyomoto}, {Makishima},
  {Diamond}, \& {Hall}}]{2001PASJ...53..215I}
{Ishihara}, Y., {Nakai}, N., {Iyomoto}, N., {Makishima}, K., {Diamond}, P., \&
  {Hall}, P. 2001, \pasj, 53, 215

\bibitem[{{Joseph} {et~al.}(2001){Joseph}, {Merritt}, {Olling}, {Valluri},
  {Bender}, {Bower}, {Danks}, {Gull}, {Hutchings}, {Kaiser}, {Maran},
  {Weistrop}, {Woodgate}, {Malumuth}, {Nelson}, {Plait}, \&
  {Lindler}}]{2001ApJ...550..668J}
{Joseph}, C.~L., {Merritt}, D., {Olling}, R., {Valluri}, M., {Bender}, R.,
  {Bower}, G., {Danks}, A., {Gull}, T., {Hutchings}, J., {Kaiser}, M.~E.,
  {Maran}, S., {Weistrop}, D., {Woodgate}, B., {Malumuth}, E., {Nelson}, C.,
  {Plait}, P., \& {Lindler}, D. 2001, \apj, 550, 668

\bibitem[{{Kondratko} {et~al.}(2005){Kondratko}, {Greenhill}, \&
  {Moran}}]{2005ApJ...618..618K}
{Kondratko}, P.~T., {Greenhill}, L.~J., \& {Moran}, J.~M. 2005, \apj, 618, 618

\bibitem[{{Kormendy} {et~al.}(1997){Kormendy}, {Bender}, {Magorrian},
  {Tremaine}, {Gebhardt}, {Richstone}, {Dressler}, {Faber}, {Grillmair}, \&
  {Lauer}}]{1997ApJ...482L.139K}
{Kormendy}, J., {Bender}, R., {Magorrian}, J., {Tremaine}, S., {Gebhardt}, K.,
  {Richstone}, D., {Dressler}, A., {Faber}, S.~M., {Grillmair}, C., \& {Lauer},
  T.~R. 1997, \apjl, 482, L139

\bibitem[{{Kormendy} \& {Richstone}(1995)}]{1995ARA&A..33..581K}
{Kormendy}, J. \& {Richstone}, D. 1995, \araa, 33, 581

\bibitem[{{Kuijken} \& {Merrifield}(1993)}]{1993MNRAS.264..712K}
{Kuijken}, K. \& {Merrifield}, M.~R. 1993, \mnras, 264, 712

\bibitem[{{Kuo} {et~al.}(2011){Kuo}, {Braatz}, {Condon}, {Impellizzeri}, {Lo},
  {Zaw}, {Schenker}, {Henkel}, {Reid}, \& {Greene}}]{2011ApJ...727...20K}
{Kuo}, C.~Y., {Braatz}, J.~A., {Condon}, J.~J., {Impellizzeri}, C.~M.~V., {Lo},
  K.~Y., {Zaw}, I., {Schenker}, M., {Henkel}, C., {Reid}, M.~J., \& {Greene},
  J.~E. 2011, \apj, 727, 20

\bibitem[{{Lauer} {et~al.}(2007){Lauer}, {Tremaine}, {Richstone}, \&
  {Faber}}]{2007ApJ...670..249L}
{Lauer}, T.~R., {Tremaine}, S., {Richstone}, D., \& {Faber}, S.~M. 2007, \apj,
  670, 249

\bibitem[{{Loyer} {et~al.}(1998){Loyer}, {Simien}, {Michard}, \&
  {Prugniel}}]{1998A&amp;A...334..805L}
{Loyer}, E., {Simien}, F., {Michard}, R., \& {Prugniel}, P. 1998, \aap, 334,
  805

\bibitem[{{Macchetto} {et~al.}(1997){Macchetto}, {Marconi}, {Axon}, {Capetti},
  {Sparks}, \& {Crane}}]{1997ApJ...489..579M}
{Macchetto}, F., {Marconi}, A., {Axon}, D.~J., {Capetti}, A., {Sparks}, W., \&
  {Crane}, P. 1997, \apj, 489, 579

\bibitem[{{Maciejewski} \& {Binney}(2001)}]{2001MNRAS.323..831M}
{Maciejewski}, W. \& {Binney}, J. 2001, \mnras, 323, 831

\bibitem[{{Macri} {et~al.}(2001){Macri}, {Stetson}, {Bothun}, {Freedman},
  {Garnavich}, {Jha}, {Madore}, \& {Richmond}}]{2001ApJ...559..243M}
{Macri}, L.~M., {Stetson}, P.~B., {Bothun}, G.~D., {Freedman}, W.~L.,
  {Garnavich}, P.~M., {Jha}, S., {Madore}, B.~F., \& {Richmond}, M.~W. 2001,
  \apj, 559, 243

\bibitem[{{Marconi} {et~al.}(2001){Marconi}, {Capetti}, {Axon}, {Koekemoer},
  {Macchetto}, \& {Schreier}}]{2001ApJ...549..915M}
{Marconi}, A., {Capetti}, A., {Axon}, D.~J., {Koekemoer}, A., {Macchetto}, D.,
  \& {Schreier}, E.~J. 2001, \apj, 549, 915

\bibitem[{{Mathur} {et~al.}(2011){Mathur}, {Fields}, {Peterson}, \&
  {Grupe}}]{2011arXiv1102.0537M}
{Mathur}, S., {Fields}, D., {Peterson}, B.~M., \& {Grupe}, D. 2011, ArXiv
  e-prints

\bibitem[{{McConnachie} {et~al.}(2005){McConnachie}, {Irwin}, {Ferguson},
  {Ibata}, {Lewis}, \& {Tanvir}}]{2005MNRAS.356..979M}
{McConnachie}, A.~W., {Irwin}, M.~J., {Ferguson}, A.~M.~N., {Ibata}, R.~A.,
  {Lewis}, G.~F., \& {Tanvir}, N. 2005, \mnras, 356, 979

\bibitem[{{Mei} {et~al.}(2007){Mei}, {Blakeslee}, {C{\^o}t{\'e}}, {Tonry},
  {West}, {Ferrarese}, {Jord{\'a}n}, {Peng}, {Anthony}, \&
  {Merritt}}]{2007ApJ...655..144M}
{Mei}, S., {Blakeslee}, J.~P., {C{\^o}t{\'e}}, P., {Tonry}, J.~L., {West},
  M.~J., {Ferrarese}, L., {Jord{\'a}n}, A., {Peng}, E.~W., {Anthony}, A., \&
  {Merritt}, D. 2007, \apj, 655, 144

\bibitem[{{Merritt}(1997)}]{1997AJ....114..228M}
{Merritt}, D. 1997, \aj, 114, 228

\bibitem[{{Merritt} {et~al.}(2001){Merritt}, {Ferrarese}, \&
  {Joseph}}]{2001Sci...293.1116M}
{Merritt}, D., {Ferrarese}, L., \& {Joseph}, C.~L. 2001, Science, 293, 1116

\bibitem[{{Miyoshi} {et~al.}(1995){Miyoshi}, {Moran}, {Herrnstein},
  {Greenhill}, {Nakai}, {Diamond}, \& {Inoue}}]{1995Natur.373..127M}
{Miyoshi}, M., {Moran}, J., {Herrnstein}, J., {Greenhill}, L., {Nakai}, N.,
  {Diamond}, P., \& {Inoue}, M. 1995, \nat, 373, 127

\bibitem[{{Neumayer}(2010)}]{2010PASA...27..449N}
{Neumayer}, N. 2010, Publications of the Astronomical Society of Australia, 27,
  449

\bibitem[{{Neumayer} {et~al.}(2007){Neumayer}, {Cappellari}, {Reunanen}, {Rix},
  {van der Werf}, {de Zeeuw}, \& {Davies}}]{2007ApJ...671.1329N}
{Neumayer}, N., {Cappellari}, M., {Reunanen}, J., {Rix}, H.-W., {van der Werf},
  P.~P., {de Zeeuw}, P.~T., \& {Davies}, R.~I. 2007, \apj, 671, 1329

\bibitem[{{Nowak} {et~al.}(2007){Nowak}, {Saglia}, {Thomas}, {Bender},
  {Pannella}, {Gebhardt}, \& {Davies}}]{2007MNRAS.379..909N}
{Nowak}, N., {Saglia}, R.~P., {Thomas}, J., {Bender}, R., {Pannella}, M.,
  {Gebhardt}, K., \& {Davies}, R.~I. 2007, \mnras, 379, 909

\bibitem[{{Onken} {et~al.}(2004){Onken}, {Ferrarese}, {Merritt}, {Peterson},
  {Pogge}, {Vestergaard}, \& {Wandel}}]{2004ApJ...615..645O}
{Onken}, C.~A., {Ferrarese}, L., {Merritt}, D., {Peterson}, B.~M., {Pogge},
  R.~W., {Vestergaard}, M., \& {Wandel}, A. 2004, \apj, 615, 645

\bibitem[{{Onken} {et~al.}(2007){Onken}, {Valluri}, {Peterson}, {Pogge},
  {Bentz}, {Ferrarese}, {Vestergaard}, {Crenshaw}, {Sergeev}, {McHardy},
  {Merritt}, {Bower}, {Heckman}, \& {Wandel}}]{2007ApJ...670..105O}
{Onken}, C.~A., {Valluri}, M., {Peterson}, B.~M., {Pogge}, R.~W., {Bentz},
  M.~C., {Ferrarese}, L., {Vestergaard}, M., {Crenshaw}, D.~M., {Sergeev},
  S.~G., {McHardy}, I.~M., {Merritt}, D., {Bower}, G.~A., {Heckman}, T.~M., \&
  {Wandel}, A. 2007, \apj, 670, 105

\bibitem[{{Paturel} {et~al.}(2003){Paturel}, {Petit}, {Prugniel}, {Theureau},
  {Rousseau}, {Brouty}, {Dubois}, \& {Cambr{\'e}sy}}]{2003A&A...412...45P}
{Paturel}, G., {Petit}, C., {Prugniel}, P., {Theureau}, G., {Rousseau}, J.,
  {Brouty}, M., {Dubois}, P., \& {Cambr{\'e}sy}, L. 2003, \aap, 412, 45

\bibitem[{{Peterson}(1993)}]{1993PASP..105..247P}
{Peterson}, B.~M. 1993, \pasp, 105, 247

\bibitem[{{Pignatelli} {et~al.}(2001){Pignatelli}, {Salucci}, \&
  {Danese}}]{2001MNRAS.320..124P}
{Pignatelli}, E., {Salucci}, P., \& {Danese}, L. 2001, \mnras, 320, 124

\bibitem[{{Pinkney} {et~al.}(2003){Pinkney}, {Gebhardt}, {Bender}, {Bower},
  {Dressler}, {Faber}, {Filippenko}, {Green}, {Ho}, {Kormendy}, {Lauer},
  {Magorrian}, {Richstone}, \& {Tremaine}}]{2003ApJ...596..903P}
{Pinkney}, J., {Gebhardt}, K., {Bender}, R., {Bower}, G., {Dressler}, A.,
  {Faber}, S.~M., {Filippenko}, A.~V., {Green}, R., {Ho}, L.~C., {Kormendy},
  J., {Lauer}, T.~R., {Magorrian}, J., {Richstone}, D., \& {Tremaine}, S. 2003,
  \apj, 596, 903

\bibitem[{{Richstone} {et~al.}(2004){Richstone}, {Gebhardt}, {Aller}, {Bender},
  {Bower}, {Dressler}, {Faber}, {Filippenko}, {Green}, {Ho}, {Kormendy},
  {Lauer}, {Magorrian}, {Pinkney}, {Siopis}, \&
  {Tremaine}}]{2004astro.ph..3257R}
{Richstone}, D., {Gebhardt}, K., {Aller}, M., {Bender}, R., {Bower}, G.,
  {Dressler}, A., {Faber}, S.~M., {Filippenko}, A.~V., {Green}, R., {Ho},
  L.~C., {Kormendy}, J., {Lauer}, T.~R., {Magorrian}, J., {Pinkney}, J.,
  {Siopis}, C., \& {Tremaine}, S. 2004, ArXiv Astrophysics e-prints

\bibitem[{{Rix} \& {White}(1992)}]{1992MNRAS.254..389R}
{Rix}, H.-W. \& {White}, S.~D.~M. 1992, \mnras, 254, 389

\bibitem[{{Robertson} {et~al.}(2006){Robertson}, {Hernquist}, {Cox}, {Di
  Matteo}, {Hopkins}, {Martini}, \& {Springel}}]{2006ApJ...641...90R}
{Robertson}, B., {Hernquist}, L., {Cox}, T.~J., {Di Matteo}, T., {Hopkins},
  P.~F., {Martini}, P., \& {Springel}, V. 2006, \apj, 641, 90

\bibitem[{{Sargent} {et~al.}(1977){Sargent}, {Schechter}, {Boksenberg}, \&
  {Shortridge}}]{1977ApJ...212..326S}
{Sargent}, W.~L.~W., {Schechter}, P.~L., {Boksenberg}, A., \& {Shortridge}, K.
  1977, \apj, 212, 326

\bibitem[{{Shapiro} {et~al.}(2006){Shapiro}, {Cappellari}, {de Zeeuw},
  {McDermid}, {Gebhardt}, {van den Bosch}, \& {Statler}}]{2006MNRAS.370..559S}
{Shapiro}, K.~L., {Cappellari}, M., {de Zeeuw}, T., {McDermid}, R.~M.,
  {Gebhardt}, K., {van den Bosch}, R.~C.~E., \& {Statler}, T.~S. 2006, \mnras,
  370, 559

\bibitem[{{Shen} \& {Gebhardt}(2010)}]{2010ApJ...711..484S}
{Shen}, J. \& {Gebhardt}, K. 2010, \apj, 711, 484

\bibitem[{{Simkin}(1974)}]{1974A&amp;A....31..129S}
{Simkin}, S.~M. 1974, \aap, 31, 129

\bibitem[{{Siopis} {et~al.}(2009){Siopis}, {Gebhardt}, {Lauer}, {Kormendy},
  {Pinkney}, {Richstone}, {Faber}, {Tremaine}, {Aller}, {Bender}, {Bower},
  {Dressler}, {Filippenko}, {Green}, {Ho}, \&
  {Magorrian}}]{2009ApJ...693..946S}
{Siopis}, C., {Gebhardt}, K., {Lauer}, T.~R., {Kormendy}, J., {Pinkney}, J.,
  {Richstone}, D., {Faber}, S.~M., {Tremaine}, S., {Aller}, M.~C., {Bender},
  R., {Bower}, G., {Dressler}, A., {Filippenko}, A.~V., {Green}, R., {Ho},
  L.~C., \& {Magorrian}, J. 2009, \apj, 693, 946

\bibitem[{{Tonry} {et~al.}(2001){Tonry}, {Dressler}, {Blakeslee}, {Ajhar},
  {Fletcher}, {Luppino}, {Metzger}, \& {Moore}}]{2001ApJ...546..681T}
{Tonry}, J.~L., {Dressler}, A., {Blakeslee}, J.~P., {Ajhar}, E.~A., {Fletcher},
  A.~B., {Luppino}, G.~A., {Metzger}, M.~R., \& {Moore}, C.~B. 2001, \apj, 546,
  681

\bibitem[{{Tran} {et~al.}(2001){Tran}, {Tsvetanov}, {Ford}, {Davies}, {Jaffe},
  {van den Bosch}, \& {Rest}}]{2001AJ....121.2928T}
{Tran}, H.~D., {Tsvetanov}, Z., {Ford}, H.~C., {Davies}, J., {Jaffe}, W., {van
  den Bosch}, F.~C., \& {Rest}, A. 2001, \aj, 121, 2928

\bibitem[{{Tremaine} {et~al.}(2002){Tremaine}, {Gebhardt}, {Bender}, {Bower},
  {Dressler}, {Faber}, {Filippenko}, {Green}, {Grillmair}, {Ho}, {Kormendy},
  {Lauer}, {Magorrian}, {Pinkney}, \& {Richstone}}]{2002ApJ...574..740T}
{Tremaine}, S., {Gebhardt}, K., {Bender}, R., {Bower}, G., {Dressler}, A.,
  {Faber}, S.~M., {Filippenko}, A.~V., {Green}, R., {Grillmair}, C., {Ho},
  L.~C., {Kormendy}, J., {Lauer}, T.~R., {Magorrian}, J., {Pinkney}, J., \&
  {Richstone}, D. 2002, \apj, 574, 740

\bibitem[{{Treu} {et~al.}(2007){Treu}, {Woo}, {Malkan}, \&
  {Blandford}}]{2007ApJ...667..117T}
{Treu}, T., {Woo}, J.-H., {Malkan}, M.~A., \& {Blandford}, R.~D. 2007, \apj,
  667, 117

\bibitem[{{Tundo} {et~al.}(2007){Tundo}, {Bernardi}, {Hyde}, {Sheth}, \&
  {Pizzella}}]{2007ApJ...663...53T}
{Tundo}, E., {Bernardi}, M., {Hyde}, J.~B., {Sheth}, R.~K., \& {Pizzella}, A.
  2007, \apj, 663, 53

\bibitem[{{Valluri} {et~al.}(2005){Valluri}, {Ferrarese}, {Merritt}, \&
  {Joseph}}]{2005ApJ...628..137V}
{Valluri}, M., {Ferrarese}, L., {Merritt}, D., \& {Joseph}, C.~L. 2005, \apj,
  628, 137

\bibitem[{{Valluri} {et~al.}(2004){Valluri}, {Merritt}, \&
  {Emsellem}}]{2004ApJ...602...66V}
{Valluri}, M., {Merritt}, D., \& {Emsellem}, E. 2004, \apj, 602, 66

\bibitem[{{van den Bosch} \& {de Zeeuw}(2010)}]{2010MNRAS.401.1770V}
{van den Bosch}, R.~C.~E. \& {de Zeeuw}, P.~T. 2010, \mnras, 401, 1770

\bibitem[{{van der Marel} {et~al.}(1998){van der Marel}, {Cretton}, {de Zeeuw},
  \& {Rix}}]{1998ApJ...493..613V}
{van der Marel}, R.~P., {Cretton}, N., {de Zeeuw}, P.~T., \& {Rix}, H.-W. 1998,
  \apj, 493, 613

\bibitem[{{van der Marel} \& {Franx}(1993)}]{1993ApJ...407..525V}
{van der Marel}, R.~P. \& {Franx}, M. 1993, \apj, 407, 525

\bibitem[{{Verolme} {et~al.}(2002){Verolme}, {Copin}, {van der Marel}, {Bacon},
  {Bureau}, {Davies}, {Miller}, \& {de Zeeuw}}]{2002MNRAS.335..517V}
{Verolme}, E.~K. et al.~{Cappellari}, M., {Copin}, Y., {van der Marel}, R.~P.,
  {Bacon}, R., {Bureau}, M., {Davies}, R.~L., {Miller}, B.~M., \& {de Zeeuw},
  P.~T. 2002, \mnras, 335, 517

\bibitem[{{Walsh} {et~al.}(2012){Walsh}, {van den Bosch}, {Barth}, \&
  {Sarzi}}]{2012ApJ...753...79W}
{Walsh}, J.~L., {van den Bosch}, R.~C.~E., {Barth}, A.~J., \& {Sarzi}, M. 2012,
  \apj, 753, 79

\bibitem[{{Walsh} {et~al.}(1997){Walsh}, {Baum}, {Malamuth}, \&
  {Goudfrooij}}]{1997stis.rept...16W}
{Walsh}, J.~R., {Baum}, S.~A., {Malamuth}, E.~M., \& {Goudfrooij}, P. 1997,
  {STIS Near-IR Fringing: Basics and Use of Contemporaneous Flats for Extended
  Sources}, Tech. rep., Space Telescope Science Institute

\bibitem[{{Wandel} {et~al.}(1999){Wandel}, {Peterson}, \&
  {Malkan}}]{1999ApJ...526..579W}
{Wandel}, A., {Peterson}, B.~M., \& {Malkan}, M.~A. 1999, \apj, 526, 579

\bibitem[{{Wiklind} \& {Henkel}(1992)}]{1992A&A...257..437W}
{Wiklind}, T. \& {Henkel}, C. 1992, \aap, 257, 437

\bibitem[{{Wyithe} \& {Loeb}(2005)}]{2005ApJ...634..910W}
{Wyithe}, J.~S.~B. \& {Loeb}, A. 2005, \apj, 634, 910

\bibitem[{{Yamauchi} {et~al.}(2012){Yamauchi}, {Nakai}, {Ishihara}, {Diamond},
  \& {Sato}}]{2012arXiv1207.6820Y}
{Yamauchi}, A., {Nakai}, N., {Ishihara}, Y., {Diamond}, P., \& {Sato}, N. 2012,
  ArXiv e-prints

\bibitem[{{Yamauchi} {et~al.}(2004){Yamauchi}, {Nakai}, {Sato}, \&
  {Diamond}}]{2004PASJ...56..605Y}
{Yamauchi}, A., {Nakai}, N., {Sato}, N., \& {Diamond}, P. 2004, \pasj, 56, 605

\end{thebibliography}

\section*{Appendix}

\begin{figure}[h]
    \mbox{\includegraphics*[height=3.7cm]{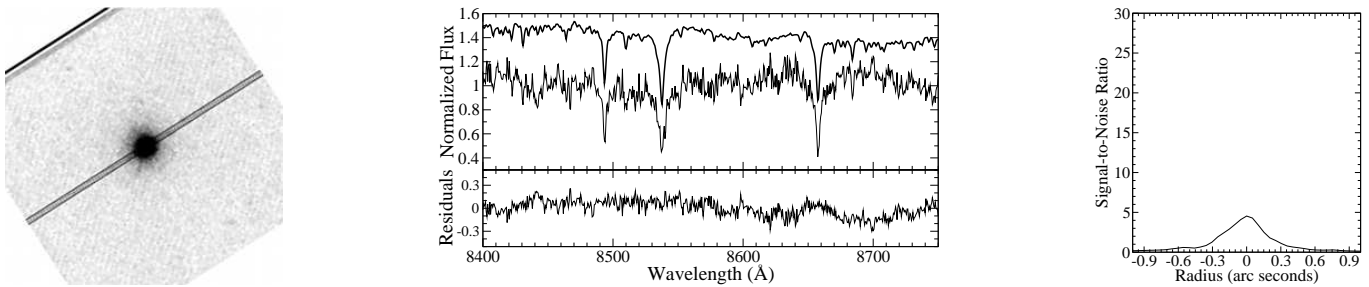}}   
    \mbox{\includegraphics*[height=3.7cm]{f6b.eps}}
    \mbox{\includegraphics*[height=3.7cm]{f6c.eps}}
\parbox{510pt}{ \caption{NGC 205. [Left] STIS acquisition image (N up, E left) and long-slit position. [Center] Peak $<$S/N$>$ spectrum with LOSVD broadened template and residuals. The LOSVD broadened template has been offset by +0.4 for clarity. [Right] $<$S/N$>$ profile.}}\label{n205}
\end{figure}
\clearpage

\begin{figure}       
     \mbox{\includegraphics*[height=3.7cm]{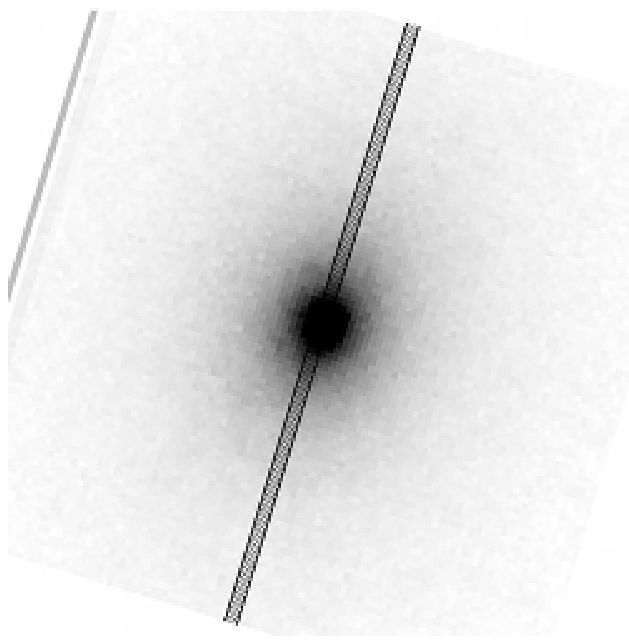}}   
    \mbox{\includegraphics*[height=3.7cm]{f7b.eps}}
    \mbox{\includegraphics*[height=3.7cm]{f7c.eps}}
\parbox{510pt}{\caption{Same as Figure 7 but for NGC 221. In this case the MPL routine recovered a double peaked LOSVD to which a second Gauss-Hermite component was fitted.}}
    \label{n221}
\end{figure}

\begin{figure}       
     \mbox{\includegraphics*[height=3.7cm]{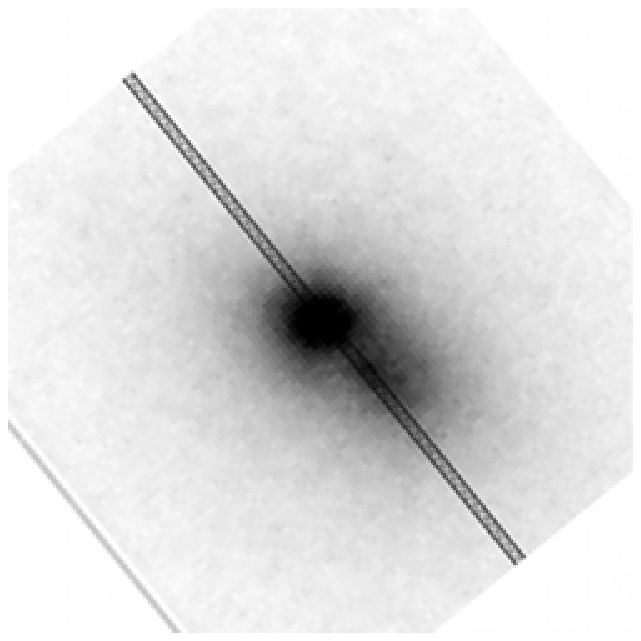}}   
    \mbox{\includegraphics*[height=3.7cm]{f8b.eps}}
    \mbox{\includegraphics*[height=3.7cm]{f8c.eps}}
\parbox{510pt}{\caption{Same as Figure~7 but for NGC 224.}}
    \label{n224}
\end{figure}

\begin{figure}       
    \mbox{\includegraphics*[height=3.7cm]{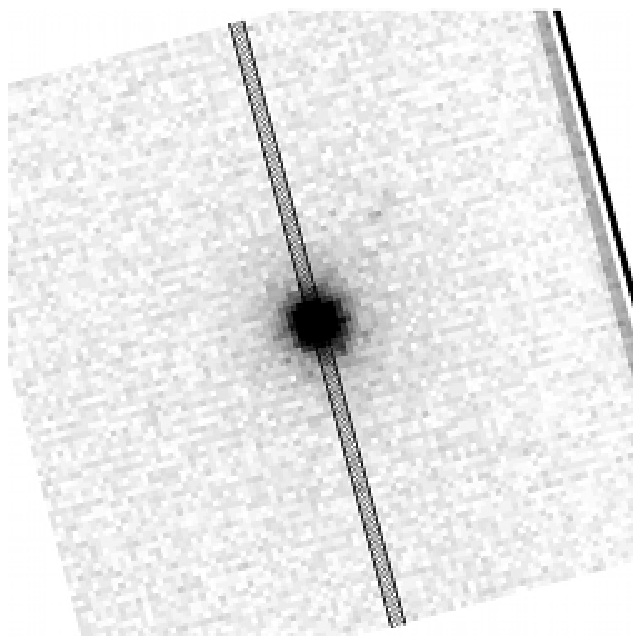}}   
    \mbox{\includegraphics*[height=3.7cm]{f9b.eps}}
    \mbox{\includegraphics*[height=3.7cm]{f9c.eps}}
\parbox{510pt}{\caption{Same as Figure~7 but for NGC 598.}}
    \label{n598}
\end{figure}

\begin{figure}      
    \mbox{\includegraphics*[height=3.7cm]{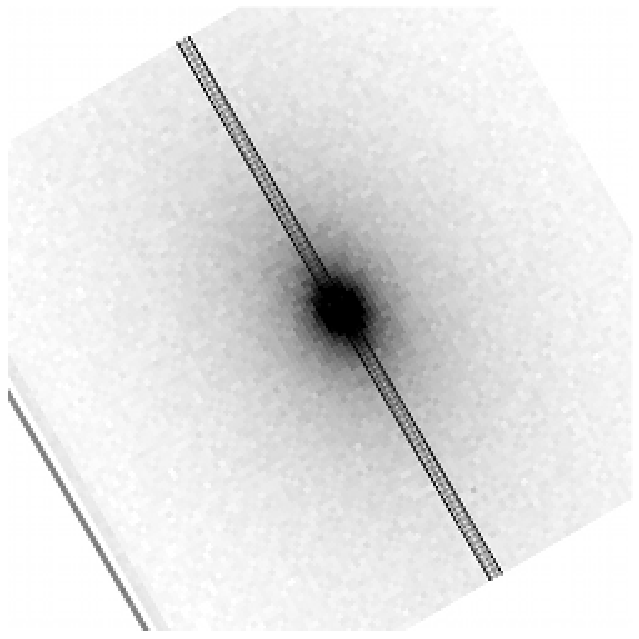}}   
    \mbox{\includegraphics*[height=3.7cm]{f10b.eps}}
    \mbox{\includegraphics*[height=3.7cm]{f10c.eps}}
\parbox[c]{510pt}{ \caption{Same as Figure~7 but for NGC 821.}}
    \label{n821}
\end{figure}

\begin{figure}
    \mbox{\includegraphics*[height=3.7cm]{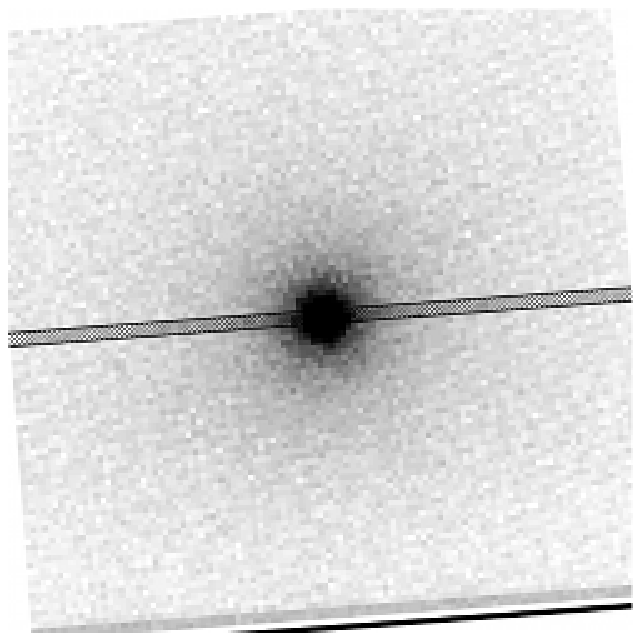}}   
    \mbox{\includegraphics*[height=3.7cm]{f11b.eps}}
    \mbox{\includegraphics*[height=3.7cm]{f11c.eps}}
\parbox{510pt}{\caption{Same as Figure~7 but for NGC 1374.}}
   \label{n1374}
\end{figure}
\clearpage

\begin{figure} 
    \mbox{\includegraphics*[height=3.7cm]{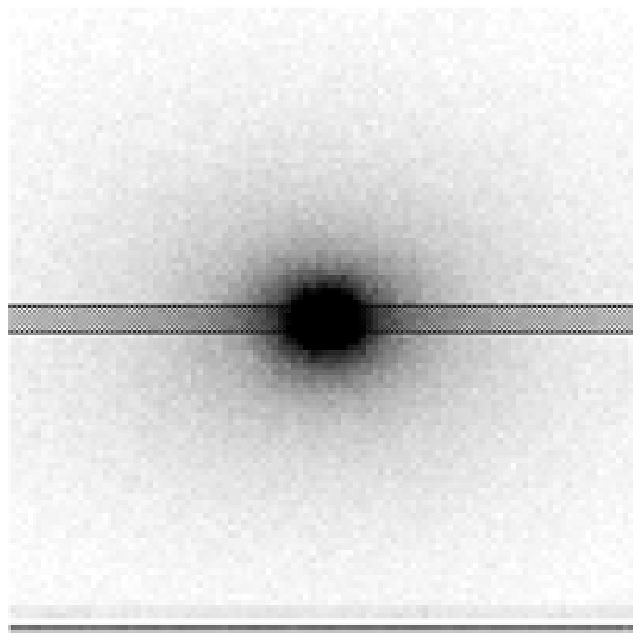}}   
    \mbox{\includegraphics*[height=3.7cm]{f12b.eps}}
    \mbox{\includegraphics*[height=3.7cm]{f12c.eps}}
\parbox{510pt}{\caption{Same as Figure~7 but for NGC 1700.}}
    \label{n1700}
\end{figure} 

\begin{figure}     
    \mbox{\includegraphics*[height=3.7cm]{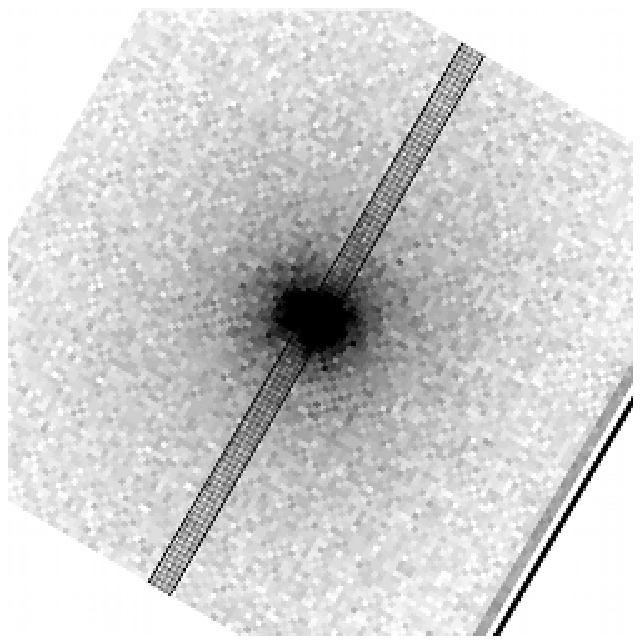}}   
    \mbox{\includegraphics*[height=3.7cm]{f13b.eps}}
    \mbox{\includegraphics*[height=3.7cm]{f13c.eps}}
\parbox{510pt}{\caption{Same as Figure~7 but for NGC 2434.}}
    \label{n2434}
\end{figure}

\begin{figure}      
    \mbox{\includegraphics*[height=3.7cm]{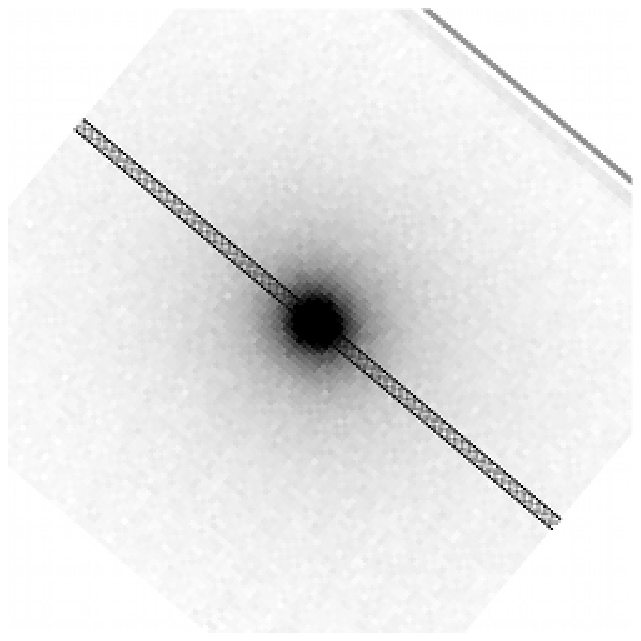}}   
    \mbox{\includegraphics*[height=3.7cm]{f14b.eps}}
    \mbox{\includegraphics*[height=3.7cm]{f14c.eps}}
\parbox{510pt}{\caption{Same as Figure~7 but for NGC 2778.}}
    \label{n2778}
\end{figure}

\begin{figure}       
    \mbox{\includegraphics*[height=3.7cm]{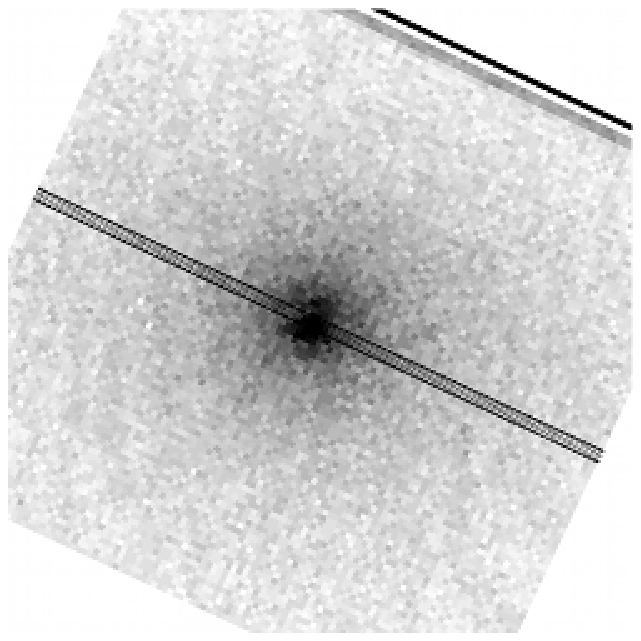}}   
    \mbox{\includegraphics*[height=3.7cm]{f15b.eps}}
    \mbox{\includegraphics*[height=3.7cm]{f15c.eps}}
\parbox[c]{510pt}{ \caption{Same as Figure~7 but for NGC 2784.}}
    \label{n2784}
\end{figure}

\begin{figure}      
    \mbox{\includegraphics*[height=3.7cm]{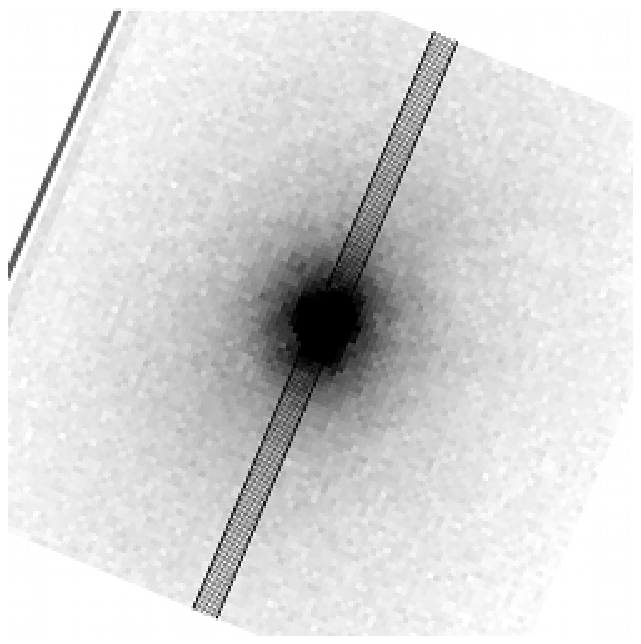}}   
    \mbox{\includegraphics*[height=3.7cm]{f16b.eps}}
    \mbox{\includegraphics*[height=3.7cm]{f16c.eps}}
\parbox{510pt}{\caption{Same as Figure~7 but for NGC 2841.}}
   \label{n2841}
\end{figure}
\clearpage

\begin{figure}     
    \mbox{\includegraphics*[height=3.7cm]{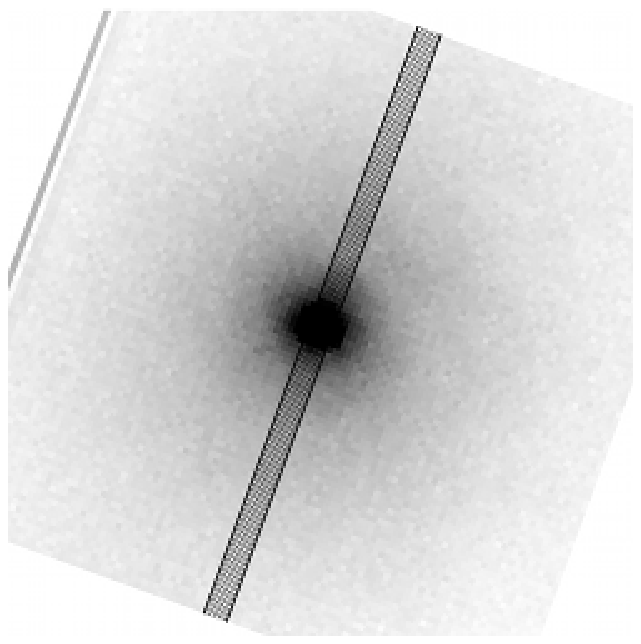}}   
    \mbox{\includegraphics*[height=3.7cm]{f17b.eps}}
    \mbox{\includegraphics*[height=3.7cm]{f17c.eps}}
\parbox{510pt}{\caption{Same as Figure~7 but for NGC 3031. No estimated LOSVD is presented for NGC 3031 due to potential contributions from AGN activity ([FeII] $8616\lambda$).}}
    \label{n3031}
\end{figure}

\begin{figure}
    \mbox{\includegraphics*[height=3.7cm]{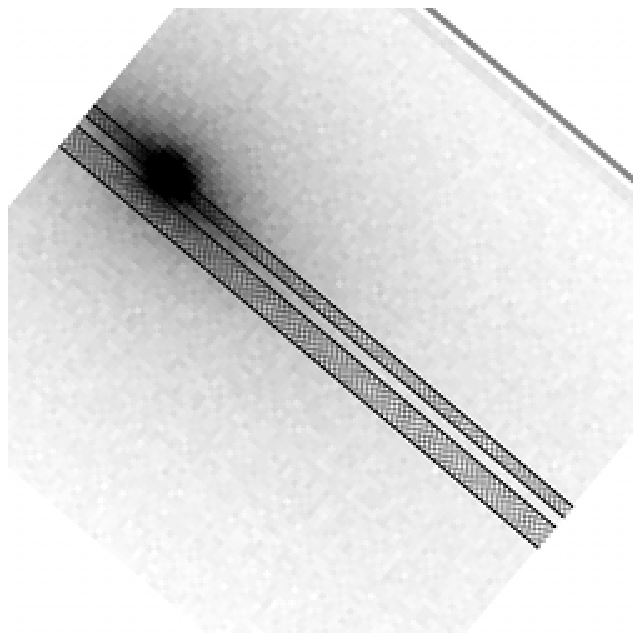}}   
    \mbox{\includegraphics*[height=3.7cm]{f18b.eps}}
    \mbox{\includegraphics*[height=3.7cm]{f18c.eps}}
\parbox{510pt}{\caption{Same as Figure~7 but for NGC 3115.}}
    \label{n3115}
 \end{figure}

\begin{figure}   
    \mbox{\includegraphics*[height=3.7cm]{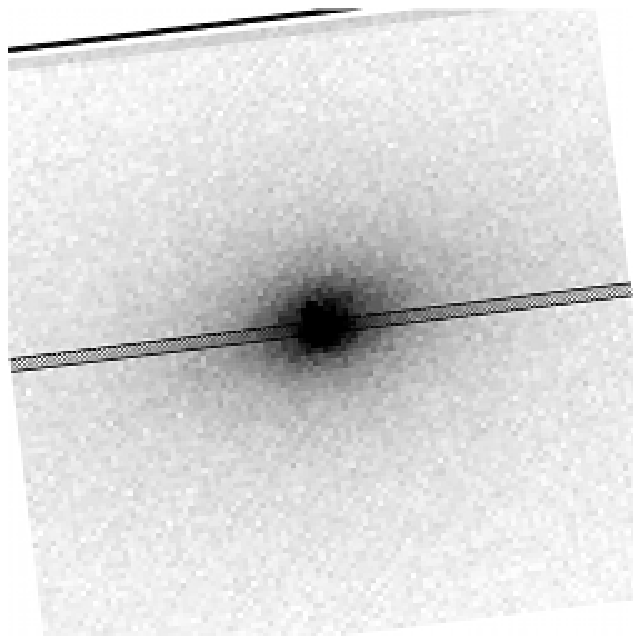}}   
    \mbox{\includegraphics*[height=3.7cm]{f19b.eps}}
    \mbox{\includegraphics*[height=3.7cm]{f19c.eps}}
\parbox{510pt}{\caption{Same as Figure~7 but for NGC 3585.}}
    \label{n3585}
\end{figure}

\begin{figure}   
    \mbox{\includegraphics*[height=3.7cm]{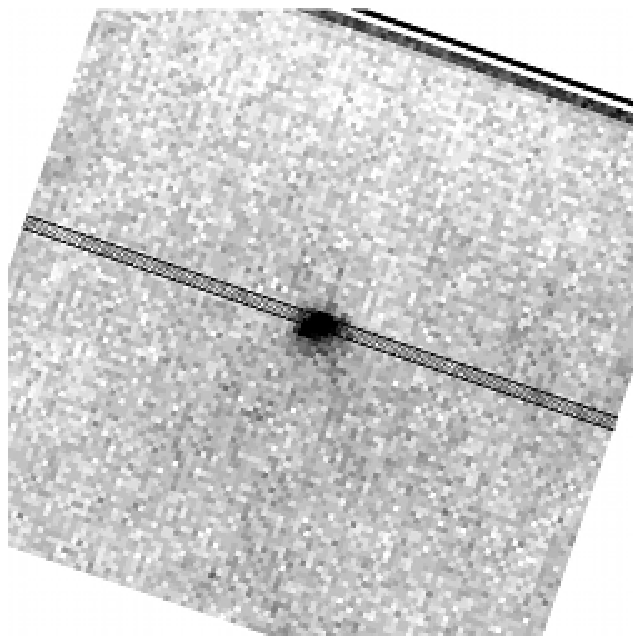}}   
    \mbox{\includegraphics*[height=3.7cm]{f20b.eps}}
    \mbox{\includegraphics*[height=3.7cm]{f20c.eps}}
\parbox{510pt}{\caption{Same as Figure~7 but for NGC 3593.}}
    \label{n3593}
\end{figure}

\begin{figure}  
    \mbox{\includegraphics*[height=3.7cm]{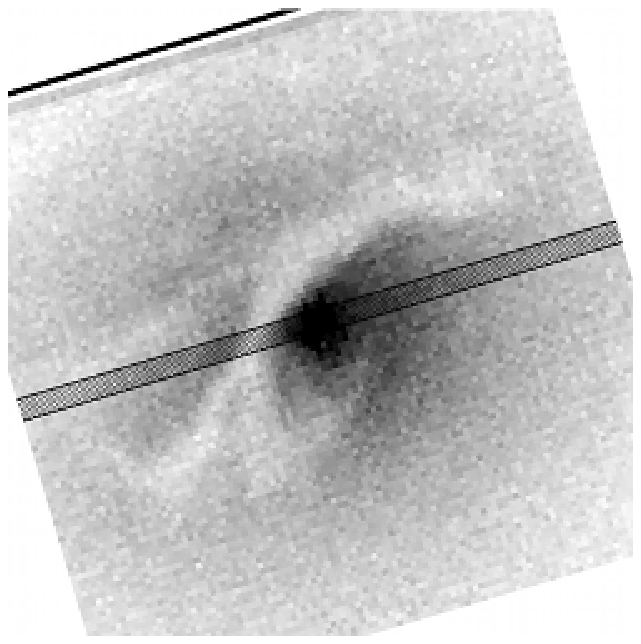}}   
    \mbox{\includegraphics*[height=3.7cm]{f21b.eps}}
    \mbox{\includegraphics*[height=3.7cm]{f21c.eps}}
\parbox{510pt}{\caption{Same as Figure~7 but for NGC 3607.}}
    \label{n3607}
\end{figure}
\clearpage

\begin{figure} 
    \mbox{\includegraphics*[height=3.7cm]{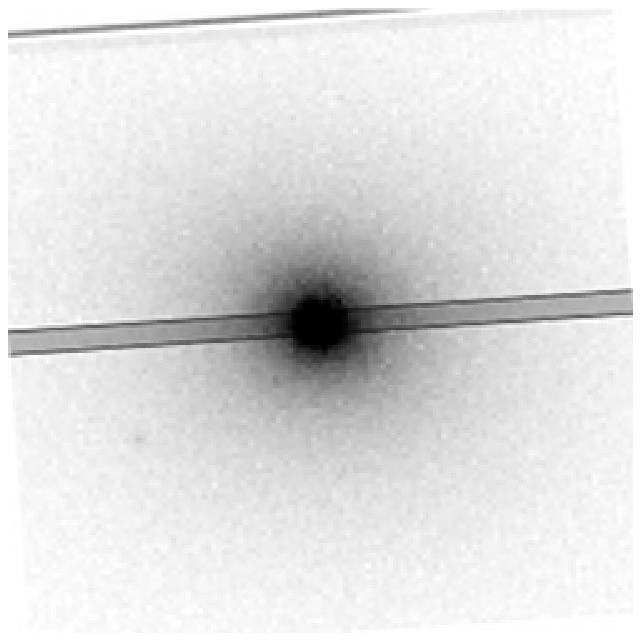}}   
    \mbox{\includegraphics*[height=3.7cm]{f22b.eps}}
    \mbox{\includegraphics*[height=3.7cm]{f22c.eps}}
\parbox{510pt}{\caption{Same as Figure~7 but for NGC 3608.}}
    \label{n3608}
 \end{figure}

\begin{figure} 
    \mbox{\includegraphics*[height=3.7cm]{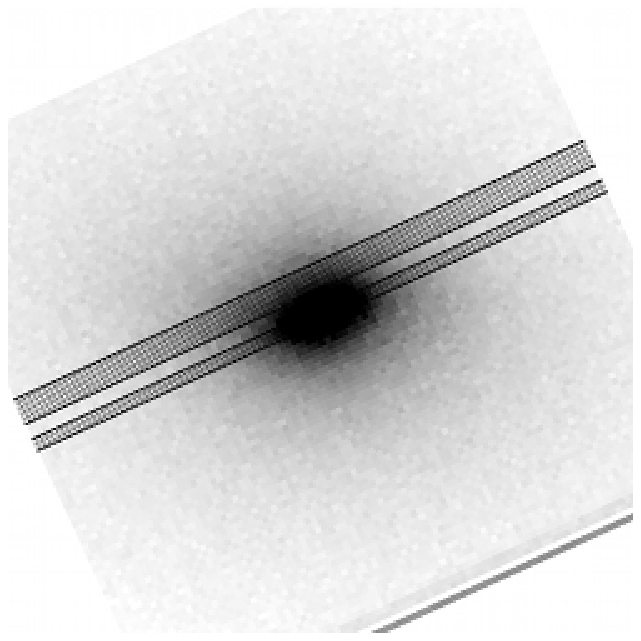}}   
    \mbox{\includegraphics*[height=3.7cm]{f23b.eps}}
    \mbox{\includegraphics*[height=3.7cm]{f23c.eps}}
\parbox{510pt}{\caption{Same as Figure~7 but for NGC 3706.}}
    \label{n3706}
 \end{figure}

\begin{figure}
    \mbox{\includegraphics*[height=3.7cm]{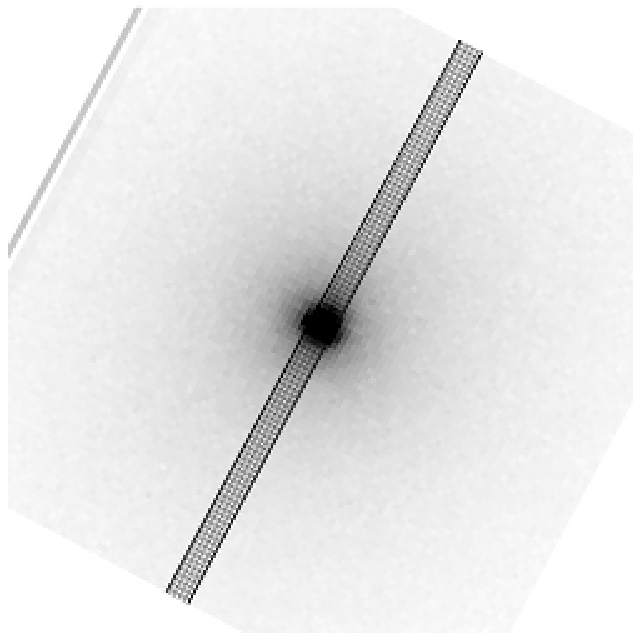}}   
    \mbox{\includegraphics*[height=3.7cm]{f24b.eps}}
    \mbox{\includegraphics*[height=3.7cm]{f24c.eps}}
\parbox{510pt}{\caption{Same as Figure~7 but for NGC 3998. No estimated LOSVD is presented for NGC 3998 due to potential contributions from AGN activity ([FeII] $8616\lambda$).}}
    \label{n3998}
\end{figure}

\begin{figure} 
    \mbox{\includegraphics*[height=3.7cm]{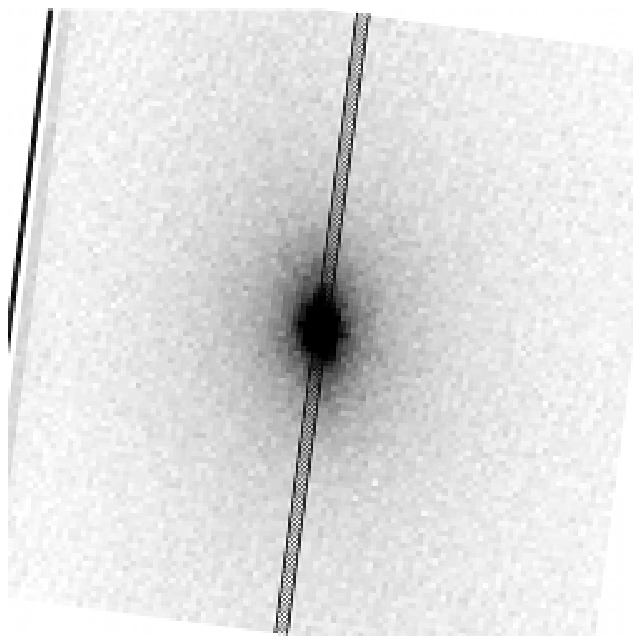}}   
    \mbox{\includegraphics*[height=3.7cm]{f25b.eps}}
    \mbox{\includegraphics*[height=3.7cm]{f25c.eps}}
\parbox{510pt}{\caption{Same as Figure~7 but for NGC 4026.}}
    \label{n4026}
 \end{figure}

\begin{figure}
    \mbox{\includegraphics*[height=3.7cm]{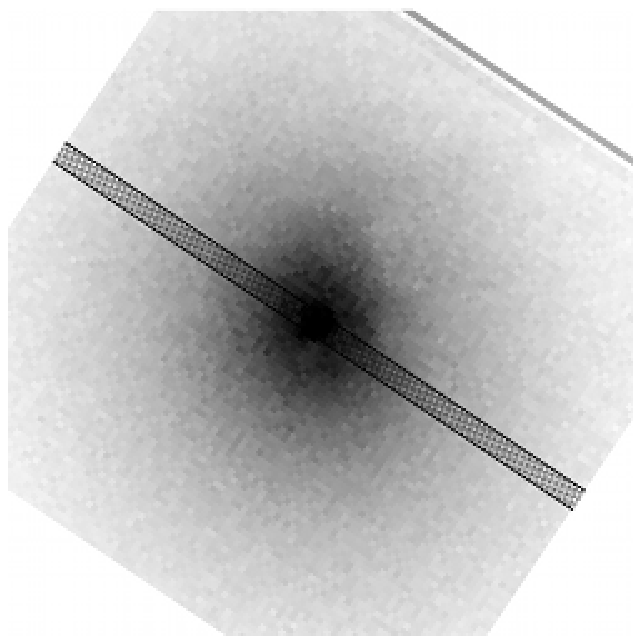}}   
    \mbox{\includegraphics*[height=3.7cm]{f26b.eps}}
    \mbox{\includegraphics*[height=3.7cm]{f26c.eps}}
\parbox{510pt}{\caption{Same as Figure~7 but for NGC 4278. No estimated LOSVD is presented for NGC 4278 due to potential contributions from AGN activity ([FeII] $8616\lambda$).}}
    \label{n4278}
\end{figure}
\clearpage

\begin{figure} 
    \mbox{\includegraphics*[height=3.7cm]{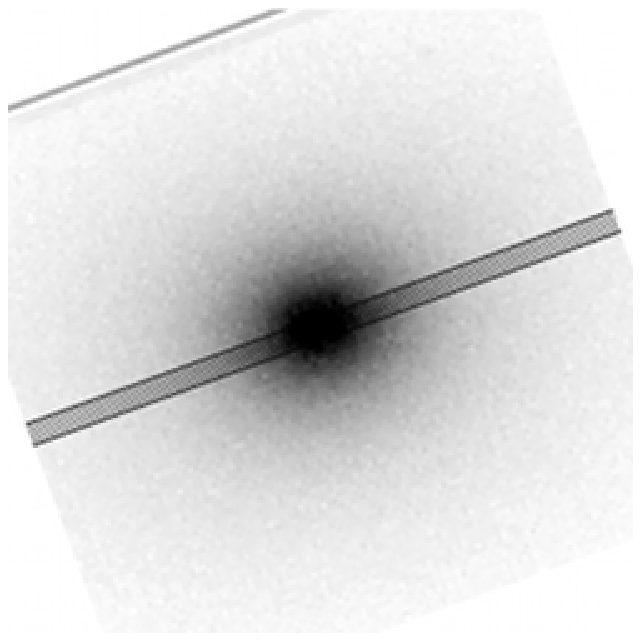}}   
    \mbox{\includegraphics*[height=3.7cm]{f27b.eps}}
    \mbox{\includegraphics*[height=3.7cm]{f27c.eps}}
\parbox{510pt}{\caption{Same as Figure~7 but for NGC 4291.}}
    \label{n4291}
 \end{figure}

\begin{figure}
    \mbox{\includegraphics*[height=3.7cm]{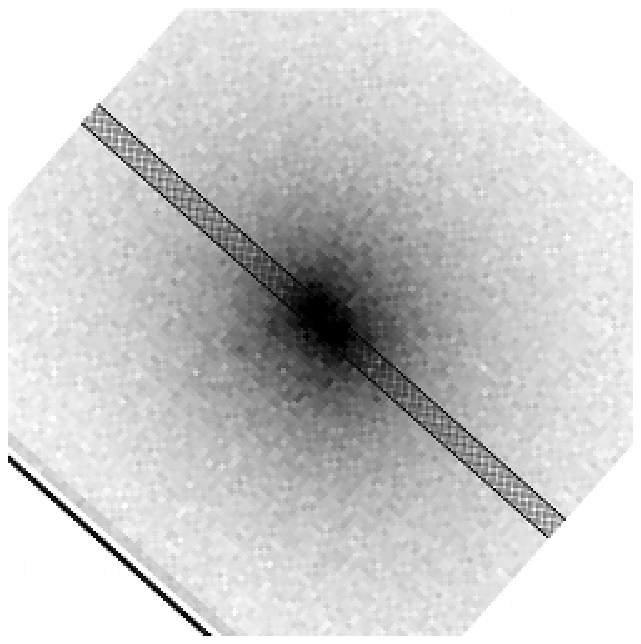}}   
    \mbox{\includegraphics*[height=3.7cm]{f28b.eps}}
    \mbox{\includegraphics*[height=3.7cm]{f28c.eps}}
\parbox{510pt}{\caption{Same as Figure~7 but for NGC 4382.}}
    \label{n4382}
\end{figure}

\begin{figure}   
    \mbox{\includegraphics*[height=3.7cm]{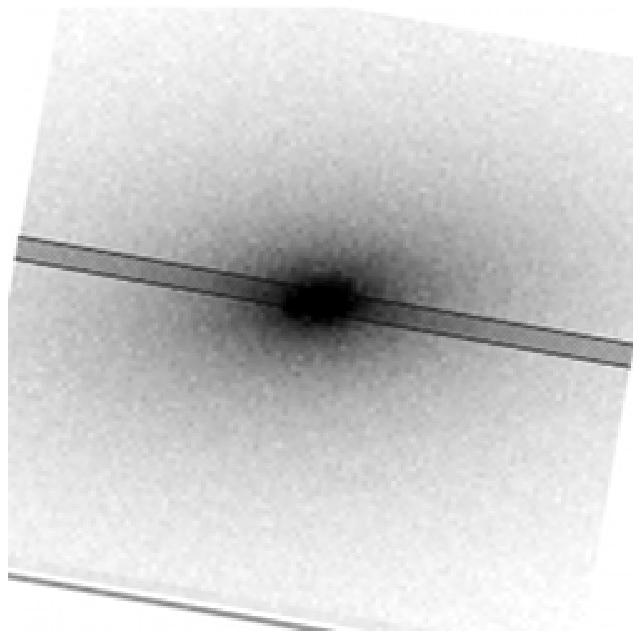}}   
    \mbox{\includegraphics*[height=3.7cm]{f29b.eps}}
    \mbox{\includegraphics*[height=3.7cm]{f29c.eps}}
\parbox{510pt}{\caption{Same as Figure~7 but for NGC 4473.}}
    \label{n4473}
 \end{figure}

\begin{figure}   
    \mbox{\includegraphics*[height=3.7cm]{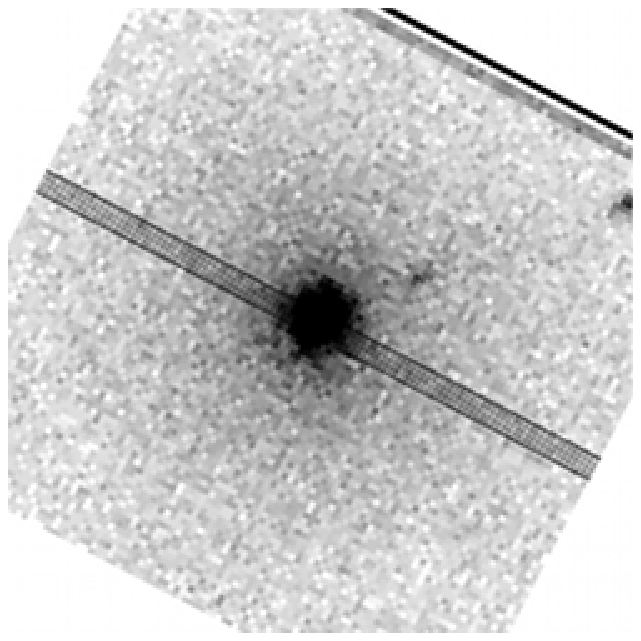}}   
    \mbox{\includegraphics*[height=3.7cm]{f30b.eps}}
    \mbox{\includegraphics*[height=3.7cm]{f30c.eps}}
\parbox{510pt}{\caption{Same as Figure~7 but for NGC 4486. No estimated LOSVD is presented for NGC 4486 due to potential contributions from AGN activity ([FeII] $8616\lambda$).}}
    \label{n4486}
\end{figure}

\begin{figure}
    \mbox{\includegraphics*[height=3.7cm]{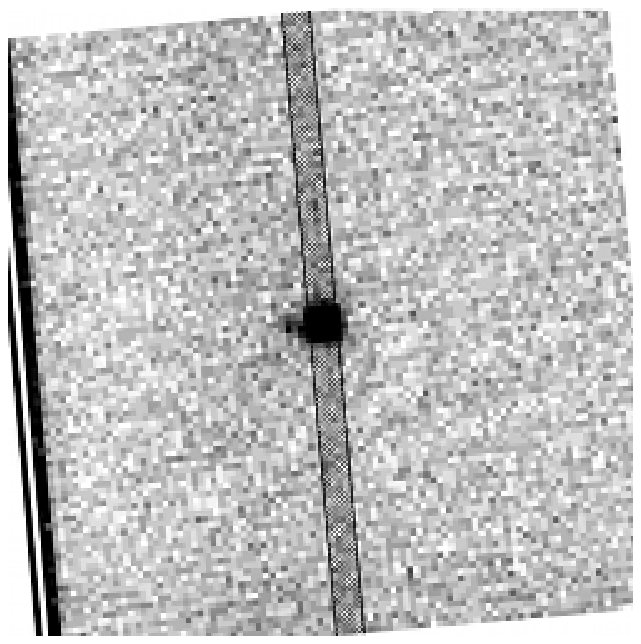}}   
    \mbox{\includegraphics*[height=3.7cm]{f31b.eps}}
    \mbox{\includegraphics*[height=3.7cm]{f31c.eps}}
\parbox{510pt}{\caption{Same as Figure~7 but for NGC 4486A.}}
    \label{n4486A}
\end{figure}
\clearpage

\begin{figure}
    \mbox{\includegraphics*[height=3.7cm]{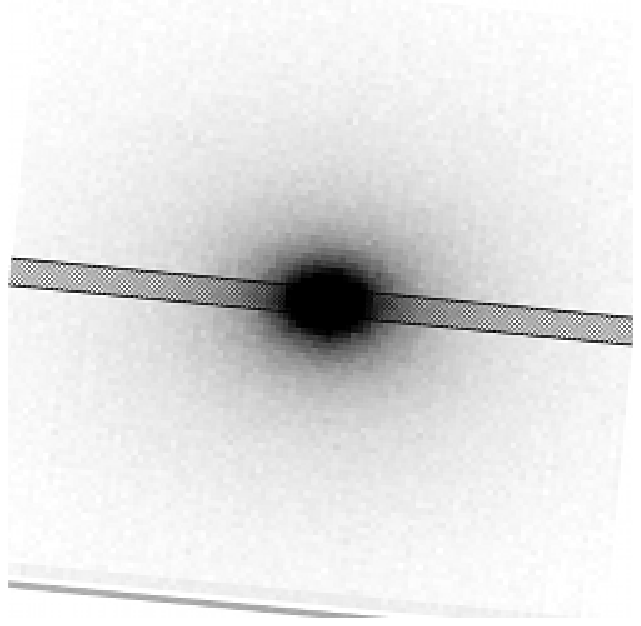}}   
    \mbox{\includegraphics*[height=3.7cm]{f32b.eps}}
    \mbox{\includegraphics*[height=3.7cm]{f32c.eps}}
\parbox{510pt}{\caption{Same as Figure~7 but for NGC 4486B.}}
    \label{n4486B}
 \end{figure}
 
\begin{figure}  
    \mbox{\includegraphics*[height=3.7cm]{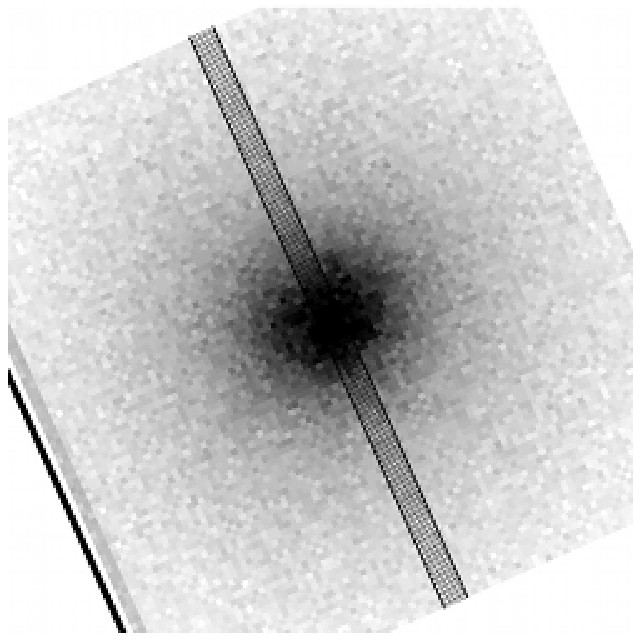}}   
    \mbox{\includegraphics*[height=3.7cm]{f33b.eps}}
    \mbox{\includegraphics*[height=3.7cm]{f33c.eps}}
\parbox{510pt}{\caption{Same as Figure~7 but for NGC 4552.}}
    \label{n4552}
 \end{figure}

\begin{figure}   
    \mbox{\includegraphics*[height=3.7cm]{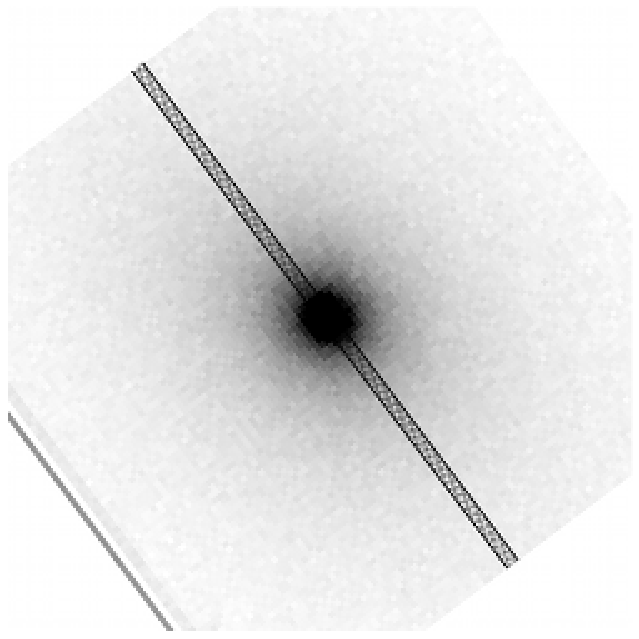}}   
    \mbox{\includegraphics*[height=3.7cm]{f34b.eps}}
    \mbox{\includegraphics*[height=3.7cm]{f34c.eps}}
\parbox{510pt}{\caption{Same as Figure~7 but for NGC 4564.}}
    \label{n4564}
\end{figure}

\begin{figure} 
    \mbox{\includegraphics*[height=3.7cm]{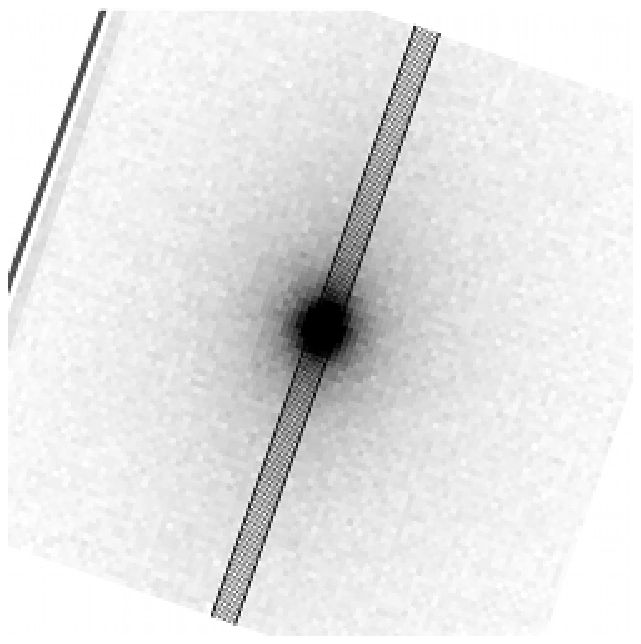}}   
    \mbox{\includegraphics*[height=3.7cm]{f35b.eps}}
    \mbox{\includegraphics*[height=3.7cm]{f35c.eps}}
\parbox{510pt}{\caption{Same as Figure~7 but for NGC 4621. No estimated LOSVD is presented for NGC 4621 due to potential contributions from AGN activity ([FeII] $8616\lambda$). }}
    \label{n4621}
 \end{figure}

\begin{figure}
    \mbox{\includegraphics*[height=3.7cm]{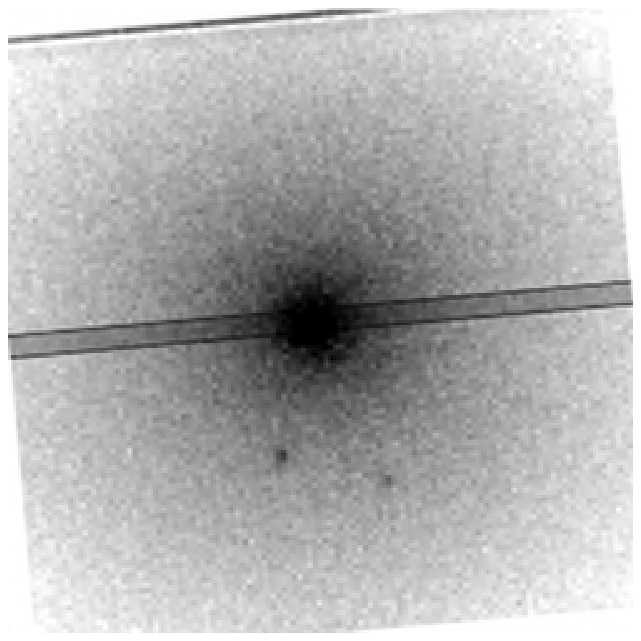}}   
    \mbox{\includegraphics*[height=3.7cm]{f36b.eps}}
    \mbox{\includegraphics*[height=3.7cm]{f36c.eps}}
\parbox{510pt}{\caption{Same as Figure~7 but for NGC 4649.}}
    \label{n4649}
 \end{figure}
\clearpage

\begin{figure}   
    \mbox{\includegraphics*[height=3.7cm]{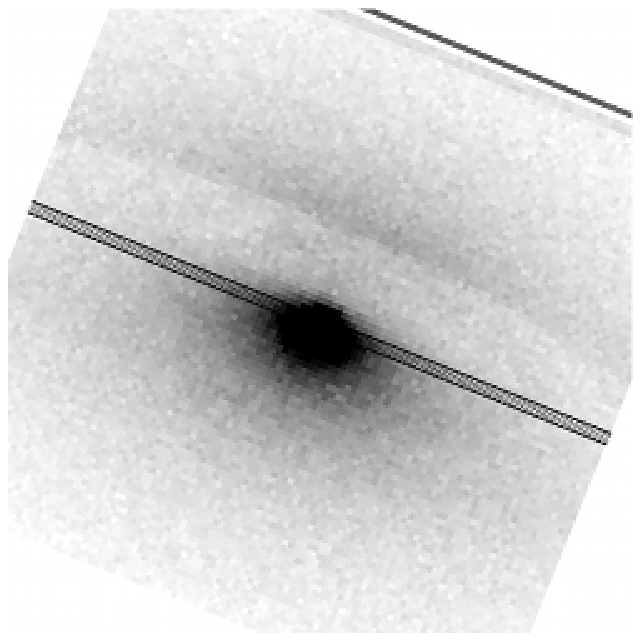}}   
    \mbox{\includegraphics*[height=3.7cm]{f37b.eps}}
    \mbox{\includegraphics*[height=3.7cm]{f37c.eps}}
\parbox{510pt}{\caption{Same as Figure~7 but for NGC 4697.}}
    \label{n4697}
 \end{figure}

\begin{figure}
    \mbox{\includegraphics*[height=3.7cm]{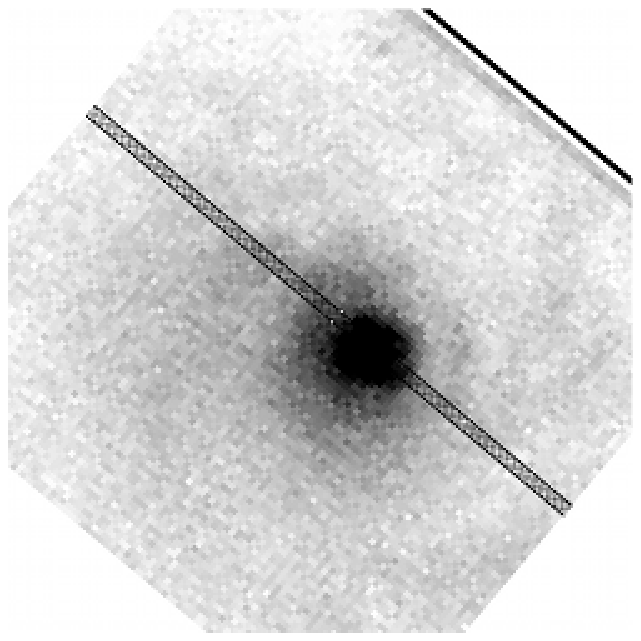}}   
    \mbox{\includegraphics*[height=3.7cm]{f38b.eps}}
    \mbox{\includegraphics*[height=3.7cm]{f38c.eps}}
\parbox{510pt}{\caption{Same as Figure~7 but for NGC 4736.}}
    \label{n4736}
 \end{figure}

\begin{figure}
    \mbox{\includegraphics*[height=3.7cm]{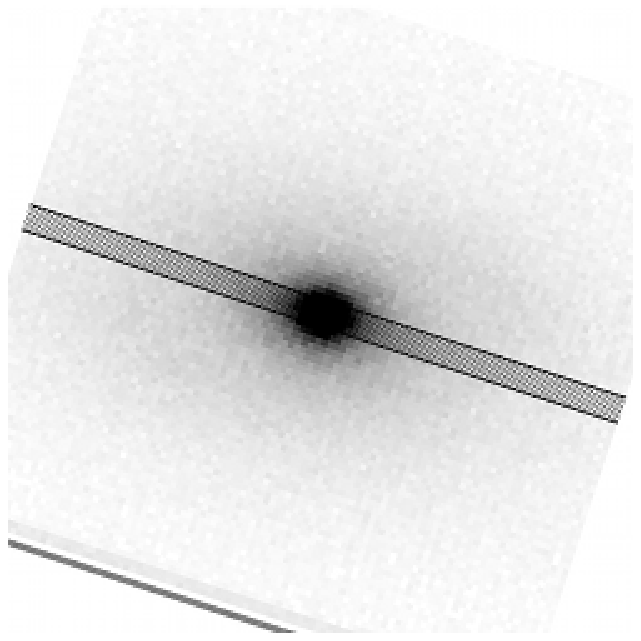}}   
    \mbox{\includegraphics*[height=3.7cm]{f39b.eps}}
    \mbox{\includegraphics*[height=3.7cm]{f39c.eps}}
\parbox{510pt}{\caption{Same as Figure~7 but for NGC 4742.}}
    \label{n4742}
 \end{figure}

\begin{figure} 
    \mbox{\includegraphics*[height=3.7cm]{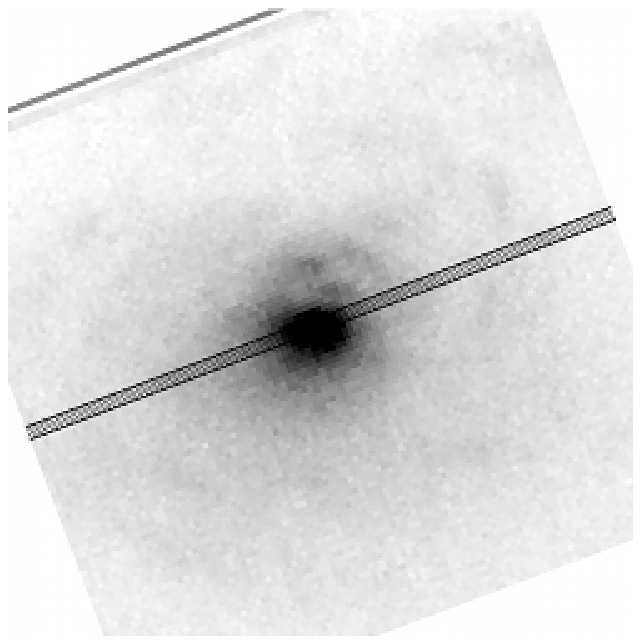}}   
    \mbox{\includegraphics*[height=3.7cm]{f40b.eps}}
    \mbox{\includegraphics*[height=3.7cm]{f40c.eps}}
\parbox{510pt}{\caption{Same as Figure~7 but for NGC 4826.}}
    \label{n4826}
 \end{figure}

\begin{figure}  
    \mbox{\includegraphics*[height=3.7cm]{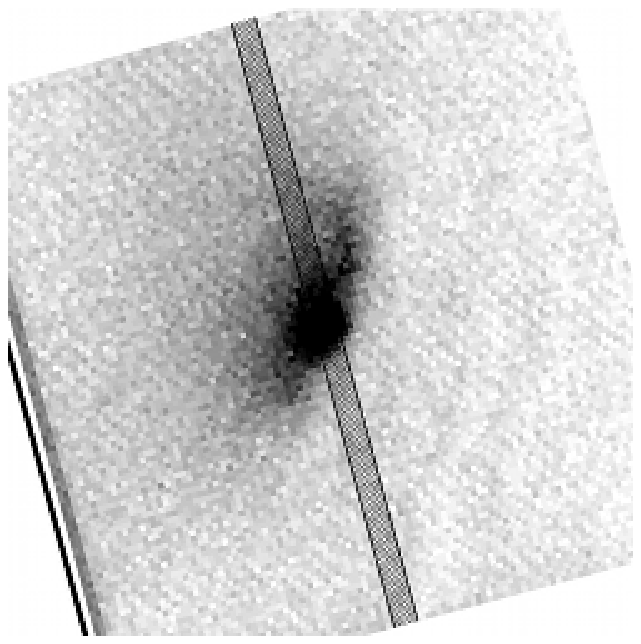}}   
    \mbox{\includegraphics*[height=3.7cm]{f41b.eps}}
    \mbox{\includegraphics*[height=3.7cm]{f41c.eps}}
\parbox{510pt}{\caption{Same as Figure~7 but for NGC 5033.}}
    \label{n5033}
 \end{figure}
\clearpage

\begin{figure}   
    \mbox{\includegraphics*[height=3.7cm]{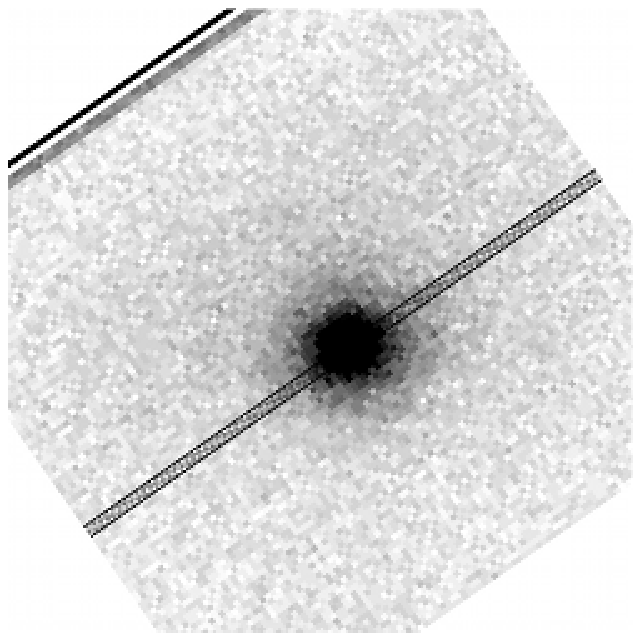}}   
    \mbox{\includegraphics*[height=3.7cm]{f42b.eps}}
    \mbox{\includegraphics*[height=3.7cm]{f42c.eps}}
\parbox{510pt}{\caption{Same as Figure~7 but for NGC 5055.}}
    \label{n5055}
 \end{figure}

\begin{figure}     
    \mbox{\includegraphics*[height=3.7cm]{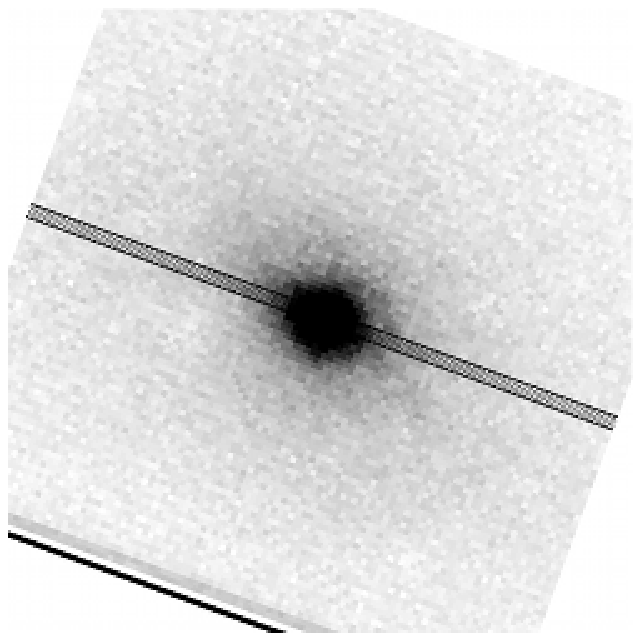}}   
    \mbox{\includegraphics*[height=3.7cm]{f43b.eps}}
    \mbox{\includegraphics*[height=3.7cm]{f43c.eps}}
\parbox{510pt}{\caption{Same as Figure~7 but for NGC 5102.}}
    \label{n5102}
 \end{figure}

\begin{figure}   
    \mbox{\includegraphics*[height=3.7cm]{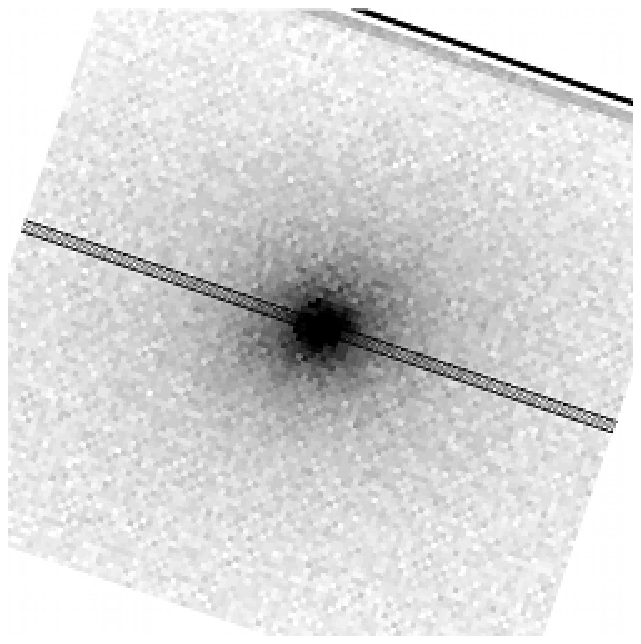}}   
    \mbox{\includegraphics*[height=3.7cm]{f44b.eps}}
    \mbox{\includegraphics*[height=3.7cm]{f44c.eps}}
\parbox{510pt}{\caption{Same as Figure~7 but for NGC 5576.}}
    \label{n5576}
 \end{figure}

\begin{figure}  
    \mbox{\includegraphics*[height=3.7cm]{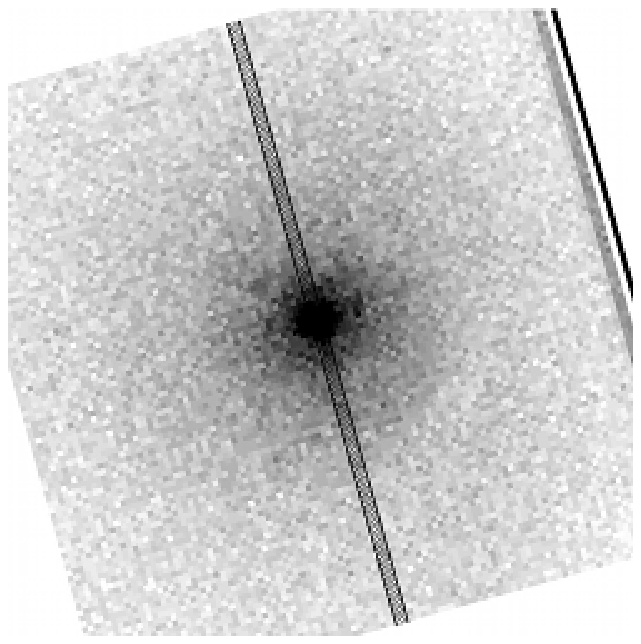}}   
    \mbox{\includegraphics*[height=3.7cm]{f45b.eps}}
    \mbox{\includegraphics*[height=3.7cm]{f45c.eps}}
\parbox{510pt}{\caption{Same as Figure~7 but for NGC 7213. No estimated LOSVD is presented for NGC 7213 due to potential contributions from AGN activity ([FeII] $8616\lambda$).}}
    \label{n7213}
\end{figure}

\begin{figure} 
    \mbox{\includegraphics*[height=3.7cm]{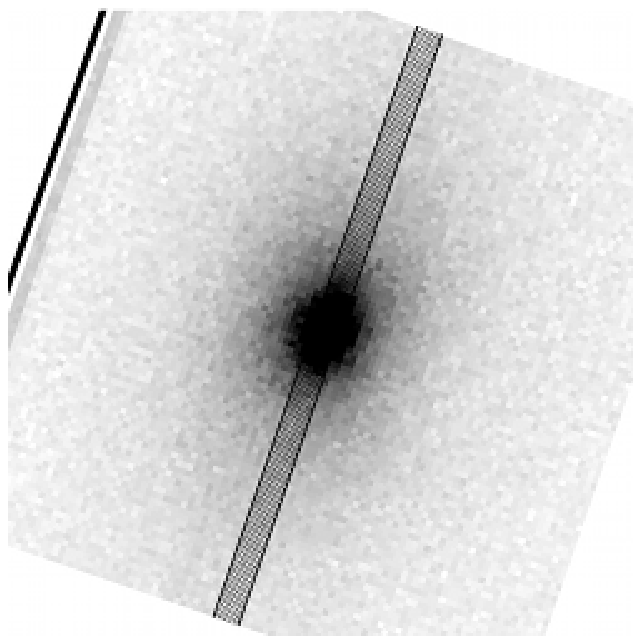}}   
    \mbox{\includegraphics*[height=3.7cm]{f46b.eps}}
    \mbox{\includegraphics*[height=3.7cm]{f46c.eps}}
\parbox{510pt}{\caption{Same as Figure~7 but for NGC 7332.}}
    \label{n7332}
 \end{figure}
\clearpage

\begin{figure} 
    \mbox{\includegraphics*[height=3.7cm]{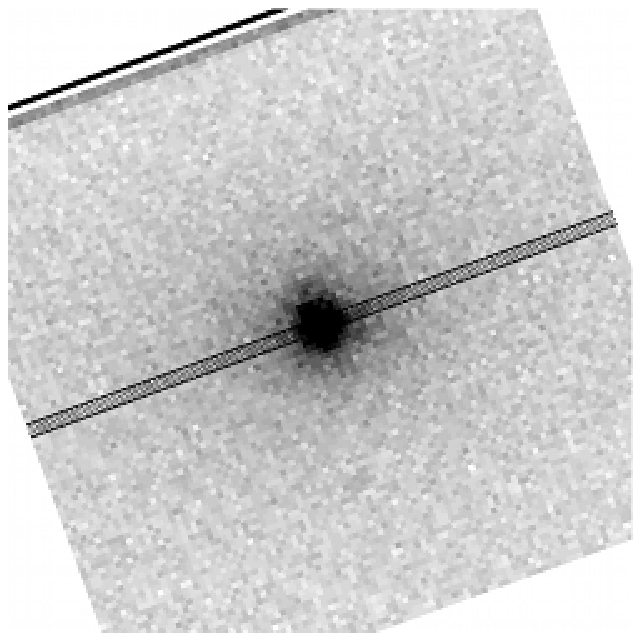}}   
    \mbox{\includegraphics*[height=3.7cm]{f47b.eps}}
    \mbox{\includegraphics*[height=3.7cm]{f47c.eps}}
\parbox{510pt}{\caption{Same as Figure~7 but for NGC 7457.  In this case the MPL routine recovered a double peaked LOSVD to which a second Gauss-Hermite component was fitted.}}
    \label{n7457}
 \end{figure}

\clearpage

\end{document}